\documentclass[amssymb,amsmath, aps,floats,preprint]{revtex4-1}
\usepackage{color,graphicx,pstricks,subfigure}
\usepackage{bm}
\usepackage[dvips]{rotating}
\usepackage{epstopdf}
\usepackage[abs]{overpic}
\usepackage{amsbsy}

\makeatletter\renewcommand\@biblabel[1]{{#1}}\makeatother

\def\r{{\mathbf{r}}}
\def\k{{\mathbf{k}}}

\def\q{{\mathbf{q}}}

\def\Tc{$T_{\rm c}$\ }
\def\i{{\rm i}}

\begin{document}

\title{\,
Disorder Induced Stripes in {\boldmath$d$}-Wave Superconductors
}
\author{{}
Markus~Schmid$^1$, Florian~Loder$^{1,2}$, Arno~P.~Kampf$^1$, and
Thilo~Kopp$^2$ \vspace{0,5cm}}

\affiliation{$^1$Center for Electronic Correlations and Magnetism, Theoretical Physics III, \\
Institute of Physics, University of Augsburg,
D-86135 Augsburg, Germany \\
$^2$Center for Electronic Correlations and Magnetism, Experimental Physics VI,\\ Institute
of Physics, University of Augsburg,
D-86135 Augsburg, Germany
}

\date{\today}

\begin{abstract}
{\bf Stripe phases are observed
experimentally in several copper-based high-${\boldsymbol T_c}$ superconductors 
near 1/8 hole doping. However, the specific characteristics may vary depending on the 
degree of dopant disorder and the presence or absence 
of a low-temperature tetragonal phase. 
On the basis of a Hartree-Fock 
decoupling scheme for the $\boldsymbol t$-$\boldsymbol J$ model we discuss the diverse behavior of stripe 
phases. 
In particular the effect of inhomogeneities is investigated in two distinctly 
different parameter regimes which are characterized by the 
strength of the interaction.  
We observe that small concentrations of impurities or vortices pin the unidirectional 
density waves, and dopant disorder 
is capable to stabilize a stripe phase in parameter regimes where homogeneous phases 
are typically favored in clean systems. 
The momentum-space results exhibit universal features 
for all coexisting density-wave solutions, nearly unchanged even in strongly 
disordered systems. These coexisting solutions feature generically 
a full energy gap and a particle-hole asymmetry in the density of states. 
}
\end{abstract}

\maketitle

\section{Introduction}

Stripe ordering phenomena on the nanoscale seem to be an inherent consequence of
electronic correlations in high-\Tc superconducting materials. They are most prominent 
close to $x = 1/8$ hole doping~\cite{tranquada:1995,kivelson:2003,tranquada:2004,fujita:2004} 
but they were also identified within a broader doping range in the pseudo-gap regime 
\cite{komiya:2002, hashimoto:2010, huecker:2011}. 
However, 
the nature of the stripe order varies significantly for different cuprate materials. 

Unidirectional charge- (CDW) and spin-density waves (SDW) have been detected 
in many cuprates by neutron scattering and x-ray experiments 
\cite{fujita:2004, tranquada:2008, dunsinger:2008,  abbamonte:2005,kim:2008,buechner:1994,
tranquada:1995,zimmermann:1998,christensen:2007,li:2007, tranquada:2004,abbamonte:2005, dunsinger:2008,
fujita:2004,sawa:2001, klauss:2000}. 
However, the 
details are strongly material dependent. 
Neutron scattering experiments on La$_{2-x-y}$Nd$_{y}$Sr$_x$CuO$_4$ (LNSCO) 
at $x=1/8$~\cite{buechner:1994,tranquada:1995,zimmermann:1998,christensen:2007} 
revealed static antiferromagnetic (AF) spin-density wave order with a period of eight 
lattice constants and a concomitant charge-density wave with half this period.  
A similar spin structure was found in La$_{2-x}$Ba$_x$CuO$_4$ 
(LBCO)~\cite{li:2007, tranquada:2004,abbamonte:2005, dunsinger:2008}, 
in La$_{2-x}$Ba$_y$Sr$_{x-y}$CuO$_4$ (LBSCO) (with $y=0.075$)
~\cite{fujita:2004}, and 
in La$_{2-x-y}$Eu$_y$Sr$_x$CuO$_4$ \cite{sawa:2001, klauss:2000}, where SDW and 
CDW coexist at and near $x=1/8$. 
A common feature of these cuprates is an anisotropic low-temperature 
tetragonal (LTT) 
phase and, in addition, a strong dopant disorder; both, the LTT structure and the dopant disorder 
are supposed to pin stripes. 
However, not all cuprates exhibit a LTT phase or dopant disorder. 
While every chemical doping of the cuprate parent compounds introduces  
disorder, its impact on the superconducting properties depend 
decisively on the distance between the CuO$_2$ planes and the dopants. 
In La$_{2-x}$Sr$_x$CuO$_4$ (LSCO) 
dopants are randomly positioned close 
to the CuO$_2$ planes generating effective disordered impurity potentials 
to the in-plane electrons. In contrast, oxygen dopants order in CuO 
chains in  YBa$_2$Cu$_3$O$_{7-\delta}$ (YBCO) that are separated from the CuO$_2$ planes by a BaO plane and 
they are roughly at twice the distance from the CuO$_2$ planes 
than Sr is in LSCO. Hence YBCO is minimally affected by the dopants' impurity potentials and 
is therefore considered as the cleanest material in the cuprate 
family.
In pure LSCO, which exhibits no LTT phase, spin stripes have been 
detected below $T_c$, 
but no CDW order~\cite{suzuki_lsco:1998,kimura:1999}.
Although no static stripes 
have so far been reported for YBCO, electron-nematic order is inferred from 
anisotropies \cite{vojta:2010} found in neutron scattering \cite{hinkov:2008} and 
thermoelectric transport \cite{daou:2010} 
measurements. 
In addition, incommensurate spin fluctuations are detected in superconducting YBCO and 
a static CDW appears in the presence of an external magnetic 
field~\cite{wu_ybco:2011,zaanen:1989}. The different 
behavior of 
YBCO may arise from both the absence of dopant disorder and the LTT structure. 
Stripe phenomena in cuprates must therefore be considered 
strongly material specific. Here we present a 
comprehensive analysis of their delicate response to the presence of disorder.
 
In the past few years the theoretical understanding of stripe formation in cuprates has 
advanced. Early on neutron-scattering data on LNSCO at hole doping 
$x=1/8$~\cite{tranquada:1995} suggested the formation of a spin-ladder structure in the CuO$_2$ 
planes. This structure is built from half-filled, three-legged spin ladders, separated by quarter 
filled chains. The AF spin structure changes sign from one ladder 
to the next, resulting in a wavelength of 
eight lattice constants. This same spin structure was later detected also in 
LBCO~\cite{tranquada:2008}. 
Indeed, theoretical models based on coupled spin-ladders describe 
\cite{vojta:2004,greiter:2010} with some success inelastic neutron scattering data on LBCO~\cite{tranquada:2004}. 
Already before the 
experimental discovery of stripes, Zaanen and Gunnarsson~\cite{zaanen:1989} and 
Machida~\cite{machida:1989} 
predicted the formation of spin stripes in doped antiferromagnets from mean-field analyses 
of the Hubbard model. The spin stripes suggested by Tranquada and coworkers 
were found also in various 
numerical calculations using the Hubbard model~\cite{fleck:2001}, the $t$-$J$ 
model~\cite{tohyama:1999, white:1998a, white:1998b, white:2000, martins:2000, hellberg:1997, 
hellberg:2000, himeda:2002} or 
the spin-fermion model \cite{buhler:2000}, and their existence is by now well established.
 
 A more delicate problem is the understanding of the coexistence of or competition between 
spin- and charge-stripe order and superconductivity. 
In disordered 
systems, $d$-wave superconductivity and antiferromagnetism can coexist~\cite{schmid:2010}. 
On the other hand, within a mean-field treatment 
of the $t$-$J$ model and variants of it, the above described spin-ladder state was found 
to coexist with a striped form of $d$-wave 
superconductivity~\cite{Raczkowski07,Yang:2009,loder:2010,loder:2011}. This superconducting 
(SC) state is modulated in space with the same period as the spin structure; its SC order 
parameter is minimal in the center of the AF spin ladder and maximal in between the spin 
stripes. Striped superconductivity in this context corresponds to a 
unidirectional pair-density wave (PDW) state 
\cite{berg_prb:2009, berg:2009}.  
Notably this PDW oscillates with twice the wavelength of 
the accompanying CDW, which is caused by periodic zero-crossings. Below we discuss 
and contrast the PDW with the periodically modulated $d$-wave superconductivity (mdSC), 
which lacks a 
sign change, thus oscillating with the same wavelength as the CDW. 
 
While the emerging spin order in these models is typically the same in a wide 
range of parameters for constant hole density, the stability of coexisting spin order and 
superconductivity varies strongly. 
The $t$-$J$ model displays two distinct limits 
which have been assessed within a mean-field approach in Ref.~\cite{loder:2011} : 
($i$) the strong $J$ limit, referred to as the so-called ``$V$-model'', where $V$ 
parametrizes the attractive nearest-neighbor interaction and 
represents the dominating contribution to the interaction (see Sec.~\ref{subsec:vmodel}) 
and ($ii$) the weak 
$J$ limit. The latter is well described by a BCS model which includes a repulsive on-site 
interaction and is appropriately named the ``$U$-model'' (see Sec.~\ref{subsec:umodel}). 
In both limits, 
spin stripes form. While in the $V$-model the spin ladders are separated by 
quasi-one dimensional (1D)~\cite{loder:2010}, mutually uncorrelated SC stripes, antiferromagnetism is 
weak in the $U$-model and coexists with two-dimensional (2D) superconductivity~\cite{loder:2011}. 
The cuprates are estimated to be placed in 
between the two limits. Some cuprate materials show characteristics which are 
qualitatively explained by 
the $V$-model, whereas others come closer to the physics of the $U$-model. 
As pointed out in Ref.~\cite{loder:2011}, for clean systems there is a balance between 
AF and SC correlations in a certain hole-doping range within which 
specific details matter.  
This regime reacts sensitively to dopant disorder and is likely to describe  
the physics of 214-compounds.

In this paper we discuss the impact of inhomogeneities on striped superconductors 
and compare the theoretical results with the experimental observations. 
In Sec.~\ref{sec:ham} we introduce the $U$- and the 
$V$-models as derived in Ref.~\cite{loder:2011} and recollect their respective mean-field theories. 
In Sec.~\ref{sec:res} we discuss the solutions of the $U$-model in the 
presence of disorder and vortices, and the effects of disorder on the 
solutions of the $V$-model. In Sec.~\ref{sec:sum} we summarize and discuss 
our results.

\section{Hamiltonian}
\label{sec:ham}

We follow the general idea that the one-band Hubbard model describes well the 
low-energy physics of the CuO$_2$ planes of the cuprates, including 
antiferromagnetism and superconductivity \cite{scalapino:1994}. 
At strong coupling, a unitary transformation maps the Hubbard model onto the 
$t$-$J$ model with an AF exchange coupling $J=4t^2/U$~\cite{macdonald:1988}. 
The non-local interaction in the $t$-$J$ model accounts for both 
superconductivity and antiferromagnetism already on the mean-field level 
\cite{kane:1989,jayaprakash:1989, fresard:1992,kagan,lee:1998}.  
Here we use an ansatz 
introduced by Kagan and Rice, in which the projection to exclude doubly occupied sites is 
replaced by an on-site repulsion term $U/2\sum_{i,\sigma}n_{i,\sigma}n_{i,-\sigma}$, 
leading back to the original $t$-$J$ model in the limit $U\rightarrow\infty$~\cite{kagan}. 
On mean-field level, this ansatz is equivalent 
to the fully decoupled BCS Hamiltonian ${\cal H}_{UV}$ with attractive nearest-neighbor 
interaction of strength $V$ and an on-site repulsion of strength $U$, if $V$ is 
identified with $J$~\cite{loder:2011}. We employ this model on a square lattice 
including  
randomly positioned on-site impurity potentials $V_i^{\rm imp}$ and a perpendicular 
orbital magnetic field.
Specifically the model Hamiltonian reads 
\begin{equation}
 \mathcal{H}_{UV} = - \sum_{ij\sigma} t_{ij} e^{\i\varphi_{ij}} c^{\dagger}_{i,\sigma} c^{}_{j,\sigma} 
- \frac{V}{2} \sum_{\langle ij \rangle,\sigma} c^{\dagger}_{i,\sigma} c^{\dagger}_{j,-\sigma} 
c^{}_{j,-\sigma} c^{}_{i,\sigma} + 
\frac{U}{2} \sum_{i,\sigma} n_{i,\sigma} n_{i,-\sigma} + \sum_{i,\sigma} \left(V_i^{\rm imp} - \mu\right) 
c^{\dagger}_{i,\sigma} c^{}_{i,\sigma},
\label{eq:ham_uv}
\end{equation}
where $c^\dagger_{i,\sigma}$ creates an electron 
on site $i$ with spin $\sigma=\ \uparrow,\downarrow$ 
and $n_{i,\sigma}=c^\dag_{i,\sigma}c_{i,\sigma}$.  
The hopping matrix
elements between nearest and next-nearest neighbor sites are denoted
by $t_{ij} = t$ and $t_{ij} = t' = -0.4\,t$, respectively. An
electron moving in the external magnetic field from site $j$ to $i$ acquires 
the Peierls phase $\varphi_{ij}=(\pi/\Phi_0)
\int^{{\bf r}_i}_{{\bf r}_j} {\bf A}({\bf r})\cdot{\rm d}{\bf r}$, where
$\Phi_0 = hc/2e$ and ${\bf
A(r)} = (0, x B)$ is the vector potential in the
Landau gauge. 
The attractive nearest-neighbor interaction is parametrized by $V>0$ and 
the chemical potential $\mu$ 
is adjusted to fix the electron density $n = \sum_i \langle n_i \rangle/N = 1 - x$,
where $x$ is the hole concentration

A mean-field decoupling of the interaction term $-V/2 
\sum_{\langle ij \rangle,\sigma} c^{\dagger}_{i\sigma} c^{\dagger}_{j-\sigma} 
c^{}_{j-\sigma} c^{}_{i\sigma}$, 
leads to the standard term for BCS type superconductivity, plus an AF 
interaction term of the form $-V\sum_{\langle ij \rangle,\sigma}\langle n_{i,\sigma}
\rangle n_{i,-\sigma}$. The latter results in an energy gain $V$ for two electrons on 
nearest-neighbor sites with antiparallel aligned spins. 
Consequently, the system exhibits an AF phase above a critical $V$ 
($V\sim t$) \cite{loder:2011}. 
In this regime double occupancies become very rare. Hence the 
term $U/2 \sum_{i\sigma} n_{i\sigma} n_{i-\sigma}$ is small compared to the $V$-term 
and has no qualitative consequences in this scenario. 
This case is represented by the $V$-model. On the other hand, if $V\ll t$, no AF order 
originates from $V$ and antiferromagnetism is controlled by $U$. This case is 
adequately depicted by the $U$-model~\cite{loder:2011}.

\subsection{${\boldsymbol U}$-model}
\label{subsec:umodel}
As discussed above, the interaction 
$-V/2\sum_{\langle ij \rangle,\sigma} c^{\dagger}_{i\sigma} c^{\dagger}_{j-\sigma} 
c^{}_{j-\sigma} c^{}_{i\sigma}$ 
is not important for magnetism in the limit $V\ll t$, but leads to the standard BCS expression 
for the SC order parameter. Thus the Hartree-Fock decoupling of $\mathcal{H}_{UV}$ 
can be restricted to the $U$-model:
\begin{align}
\mathcal{H}_U = &- \sum_{ij\sigma} t_{ij} \: e^{{\rm i} \varphi_{ij}}\: c^{\dagger}_{i\sigma}
c^{}_{j\sigma} - \mu \sum_{i\sigma} c^{\dagger}_{i\sigma} c^{}_{i\sigma}
+\sum_{\langle ij \rangle} \left(\Delta_{ij} c^{\dagger}_{i\uparrow}
c^{\dagger}_{j\downarrow} + h.c.\right)
\nonumber \\
&+ \frac{U}{2} \sum_i\left(\langle n_i \rangle n_i - \langle \sigma_i^z
\rangle \sigma_i^z \right) + \sum_{i\sigma} V_i^{\rm imp} c^{\dagger}_{i\sigma}
c^{}_{i\sigma},
\label{eq:u}
\end{align}
where $\sigma^z_i=(n_{i\uparrow}-n_{i\downarrow})/2$. 
In the following we will focus mainly on 
hole densities at and near $x=1/8$. 
The non-local pairing amplitude is defined as
\begin{equation}
 \Delta_{ij} = - V \langle c_{j\downarrow} c_{i\uparrow}\rangle.
\label{eq:def_delta}
\end{equation}
The $U$-model has been the starting point for numerous investigations on disorder- 
\cite{chen:2004, andersen:2007, harter:2007, alloul:2009,schmid:2010, andersen:2011} and 
field-induced antiferromagnetism \cite{zhu:2001, schmid:2010, schmid:2011, andersen:2011}.  
In the following we fix the pairing interaction at  $V=1.5\,t$ and 
set the strength of the non-magnetic impurity potential $V_i^{\rm imp}$ to $0.9\,t$. 
All fields, i.e. $\Delta_{ij}$,
the local electron density $\langle n_i \rangle$, and the local
magnetization $\langle \sigma_i^z
\rangle$ are calculated self-consistently from the solutions of the
associated Bogoliubov-de Gennes (BdG) equations. A detailed derivation of 
the BdG equations is presented in Refs.~\cite{schmid:2010,schmid:2011}. 
The $d$-wave order parameter on a lattice site $i$ 
is defined as
\begin{eqnarray}
 \Delta_i^d = \frac{1}{4} \left(  \Delta^d_{i,i+\hat{x}} +
       \Delta^d_{i,i-\hat{x}} - \Delta^d_{i,i+\hat{y}} - \Delta^d_{i,i-\hat{y}}\right),
\end{eqnarray}
where $\Delta^d_{i,j} = \Delta_{ij} e^{-{\rm i} \varphi_{ij}}$. 
In order to describe a possibly appearing modulation of the superconducting order 
parameter, as in the pair-density wave state 
\cite{agterberg:2008,berg:2009, vojta:2009,loder:2010, loder:2011}, we subdivide the pairing order 
parameter into 
three contributions \cite{berg:2009,vojta:2009}
\begin{equation}
 \Delta_i^d =  \Delta_0 + \Delta_{{\bf q}}\, e^{{\rm i} {\bf q}\cdot {\bf r}_i} +  \Delta_{-{\bf q}}\, 
e^{-{\rm i} {\bf q}\cdot {\bf r}_i},
\label{eq:deltaq}
\end{equation}
where the wavevector 
${\bf q}$
parametrizes the modulation of the pairing amplitude 
and reflects a finite center-of-mass momentum $\hbar\q$ of the electron pairs. 
$\Delta_0$ accounts for a homogeneous component of  
the order parameter. 
Two distinct types of SC states are possible solutions for ${\cal H}_U$: If $\Delta_0=0$, 
the SC order parameter shows a stripe pattern modulated with wave vector ${\bf q}$, 
where the superconducting stripes are separated by channels of zero pairing amplitude across 
which the SC order parameter changes sign. 
If the absolute values of  
$|\Delta_{{\bf q}}|$ 
and  $|\Delta_{-{\bf q}}|$ are smaller than the homogeneous component $|\Delta_0|$, 
$\Delta_i^d$ is always finite and modulated with 
wave vector $2{\bf q}$. Although the ``pure'' PDW with $\Delta_0=0$ was 
phenomenologically suggested to be the ground state of striped high-$T_{\rm c}$ 
superconductors~\cite{Berg09}, all calculations of mean-field type so far led to ground states 
with dominating uniform component $\Delta_0$, c.f.~Refs.~\cite{Yang:2009,loder:2011}.

\subsection{${\boldsymbol V}$-model}
\label{subsec:vmodel}

For large attractive interactions $V\sim t$ the Hartree-Fock decoupled Hamiltonian of 
Eq.~\eqref{eq:ham_uv} is effectively described by the $V$-model Hamiltonian
\begin{align}
H_V = &- \sum_{i,j,\sigma} t_{ij}  e^{\i\varphi_{ij}} c^{\dagger}_{i\sigma}
c^{}_{j\sigma} - \mu \sum_{i,\sigma} c^{\dagger}_{i\sigma} c^{}_{i\sigma}
+ \sum_{\langle i,j \rangle} \left(\Delta_{ij} c^{\dagger}_{i\uparrow}
c^{\dagger}_{j\downarrow} + h.c.\right)
\nonumber \\
&-V\sum_{\langle i,j\rangle,\sigma}\langle n_{i\sigma}\rangle 
c^\dag_{j-\sigma}c_{j-\sigma}+\sum_{i,\sigma} V_i^{\rm imp} c^{\dagger}_{i\sigma}
c^{}_{i\sigma},
\label{eq:v}
\end{align}
where $\Delta_{ij}$ is defined as in Eq.~\eqref{eq:def_delta}. 
The term $\langle n_{i\sigma}\rangle c^\dag_{j-\sigma}c_{j-\sigma}$ is responsible for 
antiferromagnetism; it is controlled by 
the same interaction parameter $V$ as the SC order parameter. 
Since in this limit 
AF correlations are strong enough to separate the system locally into AF 
regions close to half filling and into hole-rich regions far from half filling, 
the $U$-term contributes little and is therefore omitted from ${\cal H}_V$. 

For hole doping around $x=1/8$, the solutions of the BdG equations for $H_V$ in the impurity-free case 
are very similar to those of the $U$-model \cite{loder:2011}. 
The main difference is the stronger magnetization 
of the magnetic stripes 
in the $V$-model. 
Within the AF stripes, superconductivity 
vanishes, and the regions with dominant superconductivity 
are spatially separated into uncorrelated, 
quasi 1D filaments. Therefore the phase relation between them 
is not 
fixed and the ``pure'' PDW state is degenerate to a solution with finite 
$\Delta_0$. Because of the one-dimensional character of the solutions of the $V$-model, 
they barely adjust to impurities in comparison to 
the solutions of 
the $U$-model. This characteristic difference is discussed in Sec.~\ref{sec:V}.

\subsection{Momentum-space quantities}
\label{sec:mom_space_quant}
For a thorough analysis of our results we also discuss the following momentum 
space quantities. With the Fourier transform of the factorized square of the magnetization 
\begin{align}
S(\q) = \frac{1}{N^2} \sum_{ij} \langle \sigma_i^z \rangle \langle \sigma_j^z \rangle 
e^{-{\rm i} \q \cdot (\r_j - \r_i)} 
\label{eq:spinstruc_def}
\end{align}
we approximate the magnetic structure factor. 
In addition we discuss 
the absolute squares of the Fourier transformed $d$-wave order parameter 
\begin{align}
 |\Delta(\q)|^2 = \Big|\frac{1}{N}\sum_i \Delta_i^d e^{-{\rm i} \q \cdot \r_i} \Big|^2
\end{align}
and the deviation from the average electron density $n$ 
\begin{align}
 |n(\q)|^2 = \Big|\frac{1}{N} \sum_i \left(\langle n_i \rangle - n\right) e^{-{\rm i} \q \cdot \r_i} \Big|^2.
\label{eq:ft_charge}
\end{align} 
In the strong disorder limit, the suppression 
of the electron density at the random impurity sites dominates 
the charge distribution even for small impurity potentials. 
Coherent charge oscillations  
with small amplitudes  
are therefore barely visible 
in the Fourier transform $n(\q)$. 
Since we are especially interested in these coherent charge modulations  
we exclude the impurity sites in the presence of strong disorder from the sum over sites 
in the Fourier transformation. 

The momentum distribution is given by
\begin{equation}
 n(\k) = \sum_{\sigma} \langle c^{\dagger}_{\k\sigma} c^{}_{\k\sigma} \rangle,
\end{equation}
where the fermionic operators $c^{}_{\k\sigma}$ are the Fourier transform 
of the real-space operators $c^{}_{i\sigma}$. The spectral density at the 
Fermi energy $A(\k) \equiv A(\k,\omega =0)$ is 
determined from the imaginary part of the retarded Green's function
\begin{equation}
 A(\k, \omega) = -\frac{1}{\pi} \: \sum_{\sigma} {\rm Im} \langle\langle 
c^{}_{\k\sigma}; c{}_{\k\sigma} \rangle\rangle_{\omega}^{\rm ret}.
\end{equation}
$A(\k)$ allows for the determination of partially reconstructed Fermi surfaces which are 
characteristic for inhomogeneous superconductors. Homogeneous $d$-wave superconductors 
exhibit a finite gap away from the nodes on the Brillouin zone diagonals. 
$n(\k)$ is therefore continuous everywhere except across the nodes. 
However, stripe formation leads to a reconstruction of the Fermi surface attributed to 
finite pairing-momentum. In this case the pair density $P(\k)$ (Eq.~\eqref{eq:pairdens}) 
is readjusted and, 
as a result, the superconducting energy gap no longer covers all sectors of the underlying 
Fermi surface. 
Impurities, on the other hand, induce bound states within the superconducting gap, 
leading to a redistribution of spectral weight into the gap which can be observed 
in $A(\k)$. 
In this process the transition 
from partially filled to empty states in $n(\k)$
narrows, implicating a partial reconstruction of the Fermi surface.

The momentum distribution of the pair density $P(\k)$ is defined as \cite{baruch:2008,loder:2011}
\begin{equation}
P^2(\k) = \sum_{\q} \left|\langle c_{-\k+\q\downarrow} c_{\k\uparrow}\rangle\right|^2.
\label{eq:pairdens}
\end{equation} 
It measures the average correlation of the two occupied electron states 
$|{\bf k},\uparrow\rangle$ and $|-{\bf k}-{\bf q},\downarrow\rangle$ and thereby the parts 
of the Brillouin zone where electron pairing occurs.
Eventually, we discuss the information 
contained in 
\begin{equation}
 \rho_S(\k) = \frac{1}{2}\sum_{\q\sigma} \langle \sigma c^{\dagger}_{\k+\q\sigma} 
c^{}_{\k\sigma}\rangle
\end{equation}
on the spin state. 
This quantity is maximal in those parts 
of the Brillouin zone where antiferromagnetism dominates. Thus the competition between 
superconductivity and antiferromagnetism can be analyzed by comparing $P({\bf k})$ and $\rho_S({\bf k})$.

\section{Results}\label{sec:res}

\subsection{$\boldsymbol U$-model}
The Hamiltonian~\eqref{eq:u} gives rise to unidirectional stripes which have been 
identified in the real-space quantities of clean $d$-wave superconductors above a critical 
on-site repulsion $U_c$ 
\cite{chen:2002, chen:2004, loder:2011, schmid:2011} .  
For 
a clean ($V^{\rm imp} = 0$) $d$-wave superconductor ($d$SC) 
the free energy of stripe solutions is typically close to the homogeneous $d$SC solutions. 
Depending sensitively on the initial conditions of the self-consistent BdG calculations 
we find either stripes or homogeneous AF order, both coexisting 
with superconductivity. In clean systems, regular stripe solutions emerge only if 
the self-consistency loop is started from a striped initial state.
However, in the presence of perturbations, such as 
impurities or vortices, or by a rectangular lattice geometry, 
stripes are pinned and the stripe solutions become 
robust against changes of the initial conditions. 
This relates to the notion that in the 
unperturbed superconductor fluctuating 
stripes are present  
which become static by the pinning to defects 
\cite{chen:2002,chen:2004, schmid:2011}. 

In the $U$-model below a critical pairing interaction
an emerging stripe state exhibits 
a dominant uniform component of the superconducting order 
parameter, i. e. $\Delta_0 \neq 0$ in Eq.~\eqref{eq:deltaq}, in addition to the 
finite-momentum pairing amplitudes. 
In this case the superconducting order parameter $\Delta_i^d$ does not feature a sign change 
and we call this state a modulated $d$-wave supercondcutor (mdSC).
In Ref.~\cite{loder:2010b} it was shown within a momentum space formulation 
that a translation-invariant 
Hamiltonian in zero magnetic field can nevertheless 
support a  ``pure'' PDW groundstate with $\Delta_0 = 0$
for a similar set of parameters but with a stronger pairing interaction $V \gtrsim 2.2\,t$. 
There, an analytic approximation to Gor'kov's equations was considered. 
Solutions of this kind are also found in the impurity-free real-space model used here, 
although the PDW solution converges into a {\it local} energy minimum and is 
slightly higher in energy then the mdSC solution. 
Stripe solutions are indeed observed in a broad hole-doping range. 
In the strong disorder limit, however, we find stripe states only close to $x=1/8$. 

\subsubsection{Impurity-free stripe solutions}

\begin{figure}[t!]
\begin{overpic}[scale=1,unit=1mm]{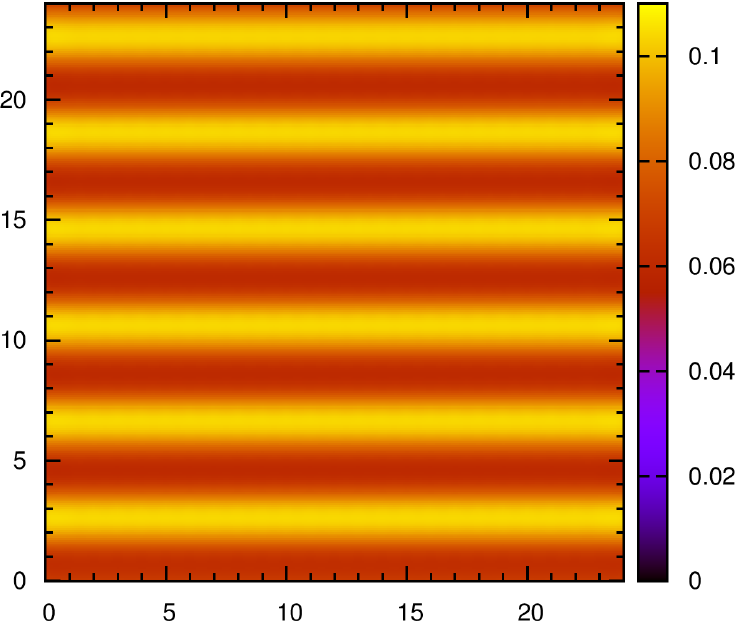}
    \put(-3,59){{\bf a}}
 \end{overpic}
\hspace{.6cm}
\begin{overpic}[scale=1,unit=1mm]{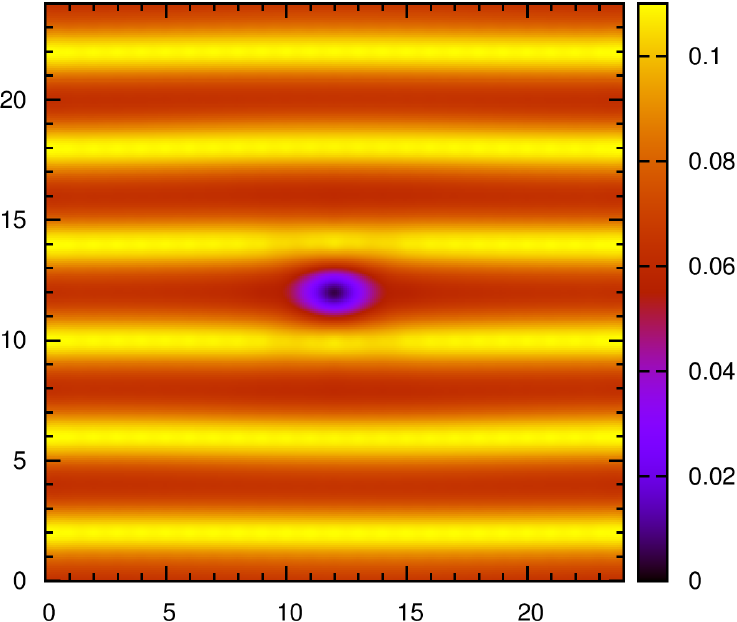}
    \put(-3,59){{\bf b}}
 \end{overpic}
\vskip6mm
\hspace{-.6cm}
\begin{overpic}[width=9cm]{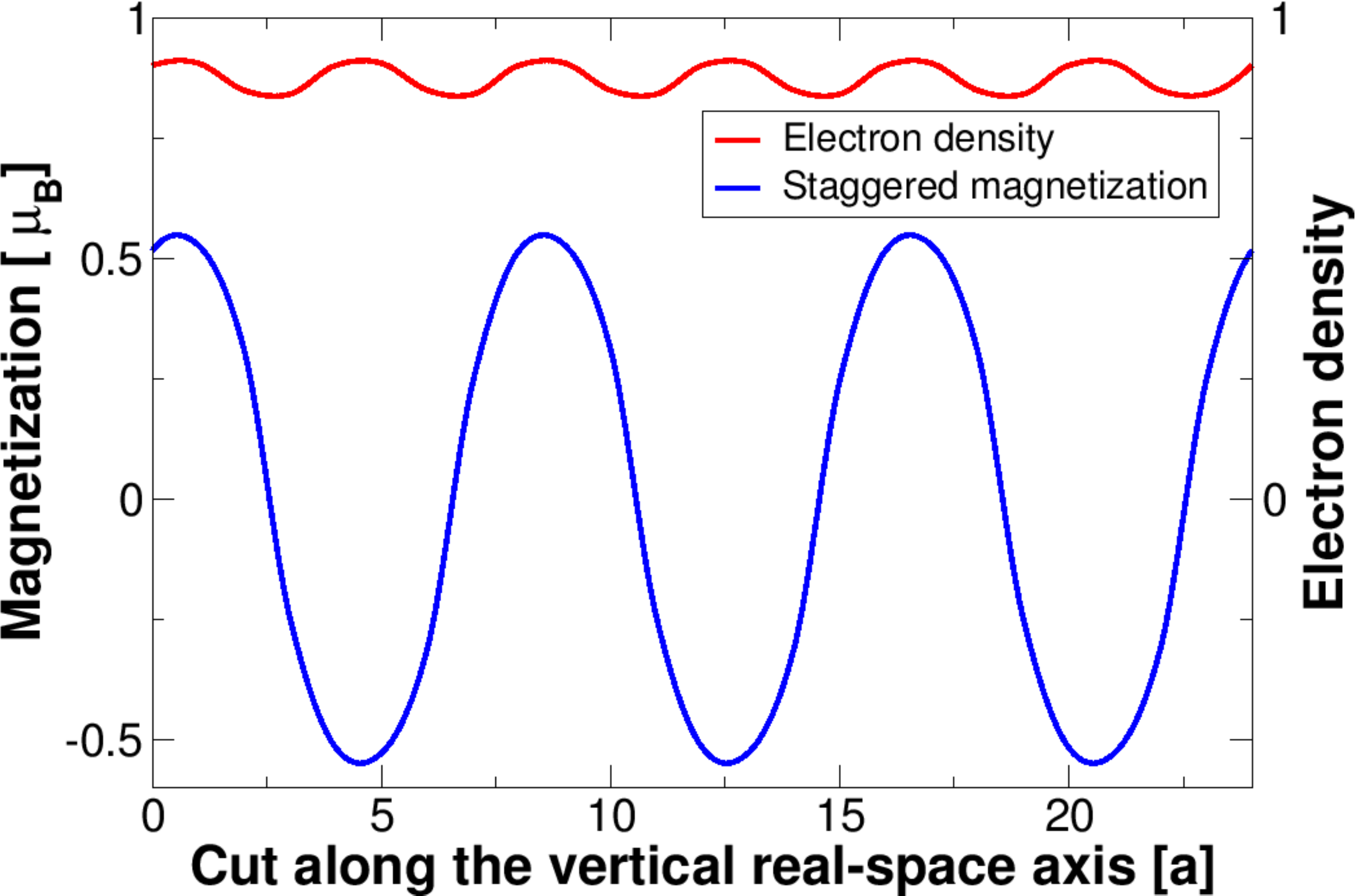}
    \put(3,157){{\bf c}}
\end{overpic}
\caption{\setlength{\baselineskip}{5mm} {\bf \boldmath Density modulations 
of a  
${\bm d}$-wave superconductor.} SC order parameter $\Delta_i^d$ for an (a) impurity-free 
and an (b) 
impurity-pinned ($V^{\rm imp} = 10t$) stripe solution. (c) Vertical cuts 
of the electron density and the staggered magnetization in the impurity-free 
case ($V^{\rm imp} = 0)$. These calculations were performed on a $24a\times 24a$ lattice 
where $a$ is the lattice constant. 
Parameters were fixed to $T=0.025\,t, t'=-0.4\,t, V = 1.5\,t, x=1/8, U=3.3t$.} 
\label{fig:impfree}
\end{figure}

Choosing sinus-shaped initial values with a wavelength of $8a$ for the self-consistent fields 
$\Delta_{ij}$ and $m_i = \langle \sigma_i^z \rangle$, 
the impurity-free system ($V^{\rm imp} = 0$) exhibits 
horizontal (or vertical) stripes in the 
real-space quantities over a wide hole-doping range. 
In Fig.~\ref{fig:impfree}
the order parameter, the electron density, 
and the staggered magnetization are shown 
for $x=1/8$. 
Obviously, a CDW, a SDW, 
and a modulated pair-density (mdSC) ($\Delta_{\q} \neq 0$) coexist in such a striped superconductor. 
The superconducting order parameter is {\it finite} everywhere, 
modulated only with an amplitude of a few percent of the average order parameter 
$\overline{\Delta}_i^d$. 
The CDW oscillates around the average density $n$ with an amplitude of only 1\%.
The staggered magnetization exhibits the strongest modulation
with an 
amplitude of about $0.5\,\mu_{\rm B}$ including a periodic sign change in the 
vertical direction corresponding to anti-phase domain walls between the antiferromagnetically 
ordered stripes. This solution is the groundstate of the $U$-model for the chosen parameters 
and initial conditions. 

The periodic modulation of the magnetization in the vertical direction translates 
into two distinct peaks in the magnetic structure factor at the wavevectors 
${\bf q}_m = (2\pi/a)\,(1/2, 1/2\pm\epsilon)$ with $\epsilon = 1/8$. 
These magnetic ordering wavevectors ${\bf q}_m$ correspond to 
a SDW with period $8a$ perpendicular to the stripes as is also obvious from the  
real space pattern of the staggered magnetization in Fig.~\ref{fig:impfree}~(c).

The concomitantly emerging 
CDW (Fig.~\ref{fig:impfree}~(c)) and mdSC (Fig.~\ref{fig:impfree}~(a))  
modulate with half the wavelength compared to the fluctuating SDW. 
Their Fourier transforms 
peak at ${\bf q}_{c/p} = 2\pi/a\,(0, \pm \delta)$ with $\delta = 1/4$ 
indicating a vertical oscillation with wavelength $\lambda_{\rm CDW/mdSC} = 4a$. 
The period doubling of the SDW
results from the $\pi$-phase shift that 
the spin order experiences 
between neighboring charge stripes. In contrast to the PDW which contains 
{\it no} uniform ${\bf q} ={\bf 0}$ pairing component, the SC order 
parameter is here always of equal sign. 
These wavelengths stay fixed 
in a hole doping range 
from $x=0.1$ up to $x=0.15$. 
For $x>1/8$, the additional holes collect in the 
already present stripes which deepens the hole rich channels thus increasing the 
oscillation amplitude of the CDW.

\begin{figure}[t!]
\hspace{.1cm}
\begin{overpic}[scale=1,unit=1mm]{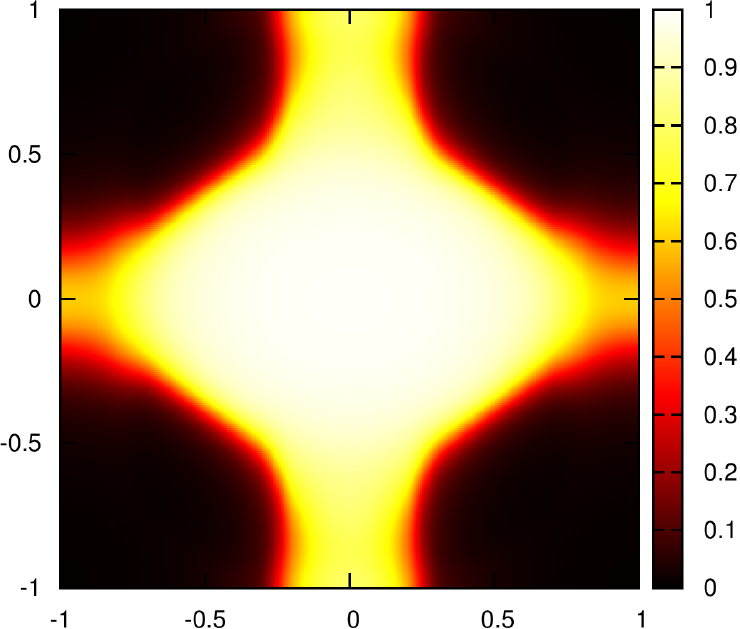}
    \put(-3,59){{\bf a}}
 \end{overpic}
\hspace{.6cm}
\begin{overpic}[scale=1,unit=1mm]{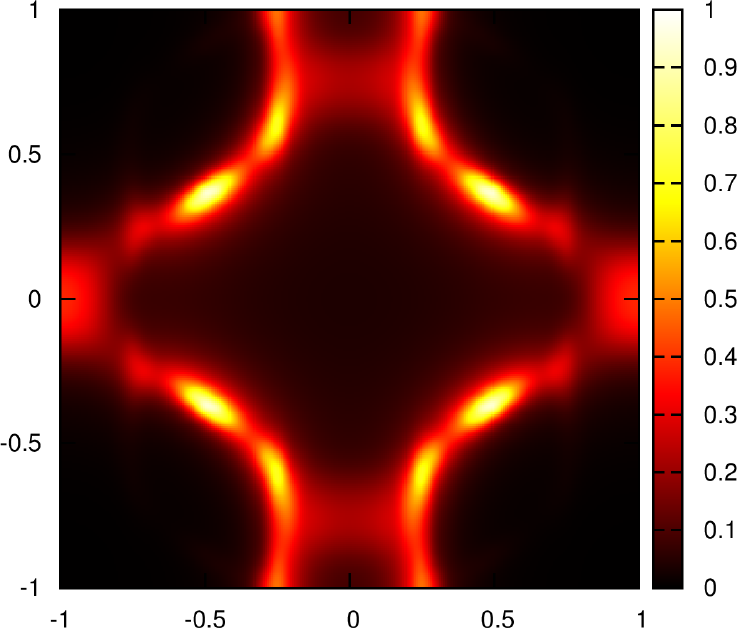}
    \put(-3,59){{\bf b}}
 \end{overpic}
\vspace{1em}
\hspace{.1cm}
 \begin{overpic}[scale=1,unit=1mm]{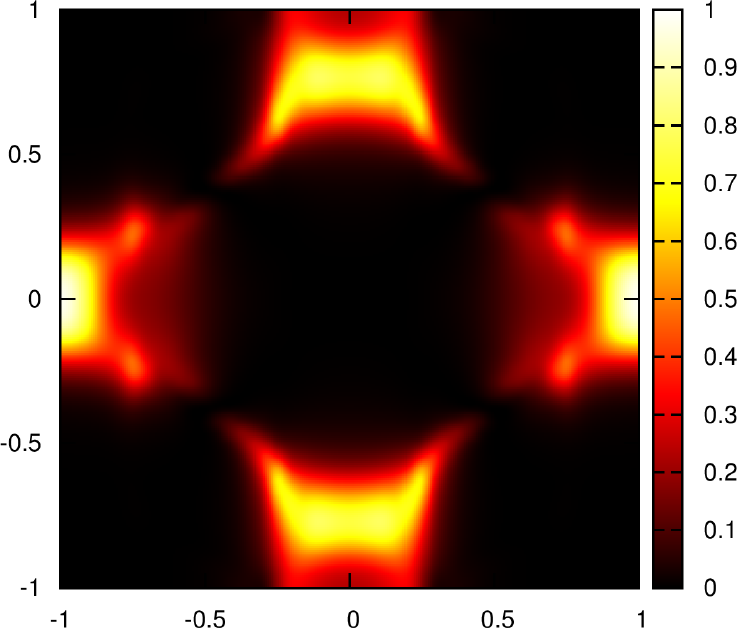}
    \put(-3,59){{\bf c}}
 \end{overpic}
\hspace{.6cm}
 \begin{overpic}[scale=1,unit=1mm]{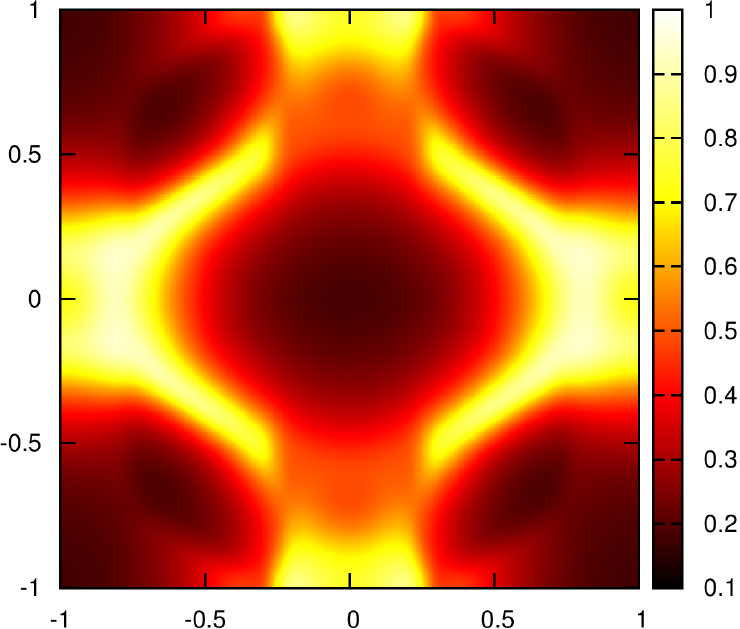}
    \put(-3,59){{\bf d}}
 \end{overpic}
\caption{\setlength{\baselineskip}{5mm} {\bf \boldmath Momentum-space 
quantities of the striped phase 
of a clean ${\bm d}$-wave superconductor.} 
a) Momentum distribution $n(\k)$, (b) spectral density $A(\k)$, 
c) pair density $P(\k)$, 
and (d) spin density $\rho_s(\k)$ for $T=0.025\,t, t'=-0.4\,t, V = 1.5\,t, x=1/8, U=3.3\,t$, 
and $V^{\rm imp} = 0$. Horizontal and vertical axes are given in units of $\pi/a$.} 
\label{fig:impfree_qspace}
\end{figure}

For a further characterization of 
the impurity-free case we present in Fig.~\ref{fig:impfree_qspace} 
momentum space quantities. 
The momentum distribution $n(\k)$ is shown in 
Fig.~\ref{fig:impfree_qspace}~(a) and 
exhibits an anisotropy in $k_x$ and $k_y$.  
In the presence of horizontal spin stripes, 
states with wavevectors near $\k \simeq (\pm\pi,0)$ are 
less probably occupied than states near $\k \simeq (0,\pm\pi)$. 

In a homogeneous $d$-wave superconductor, the spectral (quasi-particle) density $A(\k)$ is 
finite only at the four points in the Brillouin zone where the SC order parameter has nodes.
In the striped superconductor it was shown that CDW order leads to an increase of spectral 
weight in the nodal direction while SDW order suppresses spectral weight in this region 
\cite{jiang:2009}. 
Here we observe that spectral weight reappears on most of the Fermi surface of the normal conducting state (Fig.~\ref{fig:impfree_qspace}~(b)). 
In the near nodal direction two peaks
appear which are fingerprints of both 
CDW and SDW order with a dominance of the latter. 
 Comparing to Ref.~\cite{jiang:2009}, one can attribute the peak around $\k \simeq (0.25, 0.6) \pi$ in the first quadrant of the 
Brillouin zone to spin order, while the peak at $\k \simeq (0.5, 0.4)\pi$ results from 
the charge order. 
These features bear resemblance to ARPES data 
found in La$_{2-x-y}$Nd$_y$Sr$_x$CuO$_4$ and LSCO by 
Zhou {\it et al.} \cite{zhou:2001}.  
Figure~\ref{fig:impfree_qspace}~(c) shows the pair density $P(\k)$ which accentuates 
the states contributing to superconductivity. 
The maxima in $A(\k)$ around $\k \simeq (0.5, 0.4) \pi$ which originate from the 
CDW order \cite{jiang:2009} 
correspond to a suppression of the pairing in Fig.~\ref{fig:impfree_qspace}~(c). 
$P(\k)$ is maximal around $\k \simeq (\pm \pi, 0)$ where the pairing amplitude 
is largest. These regions resemble 
those of a homogeneous $d$SC but stretching 
out less far 
into the nodal direction. In contrast, $P(\k)$ is strongly suppressed at 
wavevectors $\k \simeq (0, \pm \pi)$ where the spin density is strong 
and exhibits local maxima around $\k \simeq (0, \pm 3/4 \pi)$.  
The unidirectional character of the spin-density wave 
is clearly seen in $\rho_s(\k)$ (see Fig.~\ref{fig:impfree_qspace}~(d)) 
which relates to the large amplitude oscillations of the 
staggered magnetization in real space.
$\rho_s(\k)$ dominates regions in momentum space which $P(\k)$ would occupy in a non-magnetic 
homogeneous $d$SC. It extends far along the Fermi arcs, crossing 
the nodal points. 
$\rho_s(\k)$ has large intensities in the anti-nodal regions, i.e., it competes with superconductivity.
Though maxima of $\rho_s(\k)$ 
are clearly separated from those of the pair density $P(\k)$, both quantities coexist in 
large parts of momentum space.

\subsubsection{Impurity-pinned stripes}

In the presence of a {\it single} strong impurity with potential strength $V^{\rm imp} = 10\,t$, 
stripes are pinned as is inferred from the order parameter pattern shown in 
Fig.~\ref{fig:impfree}~(b), where the impurity is located at the center.
Impurity-pinned stripe solutions are robust against a change of the initial conditions, 
in contrast to the 
impurity-free 
stripe solution ((Fig.~\ref{fig:impfree}~(a)). Thus we expect fluctuating stripes in real materials 
to be pinned by 
impurities and to become static. 
Horizontally aligned stripes emerge as well in the magnetization and the electron density. 
The impurity pins a channel of 
reduced pairing-amplitude (Fig.~\ref{fig:impfree}~(b)), which minimizes the loss of 
pairing energy. 
On the other hand, inhomogeneities in the order parameter result in a charge density 
redistribution in the absence of particle-hole symmetry~\cite{schmid:2011}. 
In fact, the channels of reduced pairing amplitude collect electrons, thereby 
shifting the electron density in these channels towards half-filling. 
This in turn favors the emergence of antiferromagnetic order in the regions of enhanced electron 
density. 
That is why the impurity pins one of the ridges of the staggered 
magnetization. 
Altogether, electron-rich stripes coincide with strongly magnetized stripes 
of reduced pairing amplitude. 
The characteristics of the impurity-pinned stripes are similar 
to those of the  impurity-free case, except that the impurity-pinned stripes emerge 
independently of the initial conditions. 
Thus the momentum space quantities shown in Fig.~\ref{fig:impfree_qspace} for the impurity-free 
stripe solution are essentially the same for the impurity-pinned stripes. 
Summarizing, we observe for sufficiently large on-site repulsion $U > U_c$, 
as in the impurity-free case, 
the coexistence of a mdSC, a SDW, and a CDW with wavelength 
 $\lambda_{\rm CDW} = \lambda_{\rm mdSC} = 4a = 1/2\,\lambda_{\rm SDW}$ 
pinned by a single non-magnetic impurity. 

Real-space quantities of stripes pinned by a single strong non-magnetic impurity have already been 
investigated by Chen and Ting \cite{chen:2003}. They obtained similar results for the $U$-model 
at $x=15\%$ hole doping which is assumed to be close to optimal 
doping.
We also checked that the periodicity of the stripes for a single impurity at $x=1/10$ 
remains the same as for $x = 1/8$.

\subsubsection{Vortex-pinned stripes}

\begin{figure}[t!]
\centering
\hspace{.1cm}
\begin{overpic}[scale=1,unit=1mm]{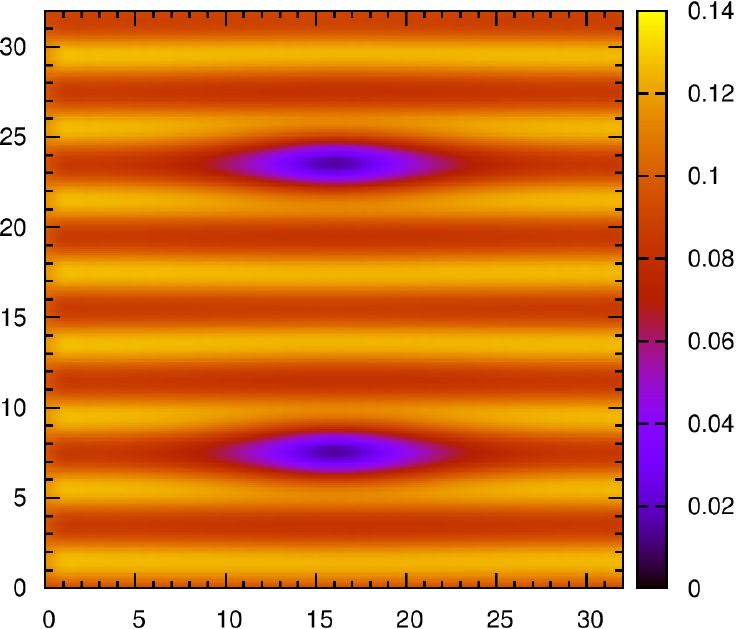}
    \put(-3,59){{\bf a}}
\end{overpic}
\hspace{.6em}
\begin{overpic}[scale=1,unit=1mm]{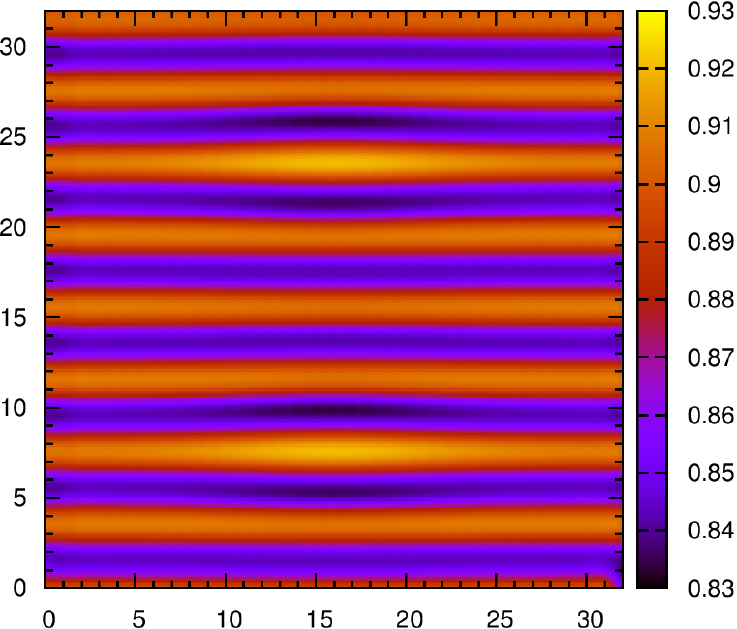}
    \put(-3,59){{\bf b}}
 \end{overpic}
\vskip7mm
\begin{overpic}[scale=1,unit=1mm]{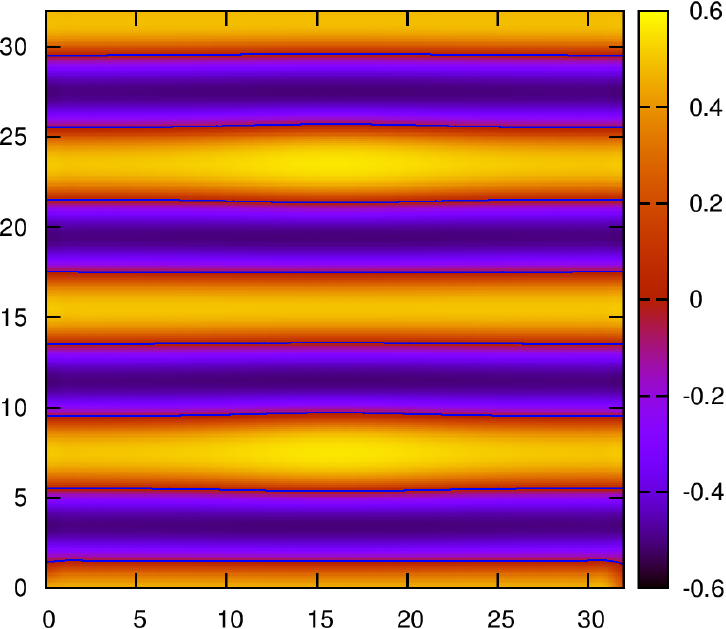}
    \put(-3,59){{\bf c}}
 \end{overpic}
\caption{\setlength{\baselineskip}{5mm}{\bf Vortex-pinned stripes.} Results are shown for a magnetic 
flux $\Phi = 2\Phi_0$ penetrating a $32a\times32a$ lattice which corresponds to a magnetic field $B = 25.6$\,T. 
(a) SC order parameter, (b) electron density, (c) staggered magnetization (blue lines mark the zero-crossing). 
The model parameters were set to 
$T=0.025\,t, t'=-0.4\,t, V = 1.34\,t, x=1/8, U=3.2\,t$.} 
\label{fig:stripes_u3_field}
\end{figure}

The vortex-pinned stripes shown in Fig.~\ref{fig:stripes_u3_field} display essentially the same 
properties as those for the impurity-free and the impurity-pinned stripes. 
The amplitudes of the 
mdSC, CDW, and the SDW are marginally enhanced as compared to the impurity-pinned stripes 
because a
vortex penetrating the superconductor 
acts as a far stronger perturbation
than a single pointlike impurity. 
Just like the impurity, the vortices pin the ridges of the SDW. 
Stripes of reduced pairing run through the elliptically deformed vortices and thereby 
save condensation energy. 
The AF stripes are further attracted by the vortex since vortex cores in
$d$-wave superconductors strongly magnetize 
due to a spin-dependent splitting of the Andreev bound-state~\cite{schmid:2010, schmid:2011}. 
Correspondingly, the vortex pins an electron-rich stripe with a 
filling shifted towards 1/2. 
From the Fourier transforms of the real-space quantities we obtain 
the same wavelength of the mdSC ($\lambda_{\rm mdSC}=4a$), 
the CDW ($\lambda_\text{CDW} = 4a$), 
and the SDW ($\lambda_\text{SDW} = 8a$) at $x=1/10$ and at $x=1/8$. 
Similar results for coexisting CDWs and SDWs were observed in Ref. \cite{chen:2002} 
within the $U$-model 
for vortex-pinned stripes 
above a critical $U$ at $x=15\%$ hole doping. 
Hence the wavelengths of these unidirectional density waves are identical in 
the impurity-free and the impurity- and vortex-pinned stripes and robust against 
a change of hole doping or pairing interaction strength, as we verified 
by explicit calculations. 
 
\subsubsection{Disorder-pinned stripes}

\begin{figure}[p!]
\begin{overpic}[scale=1,unit=1mm]{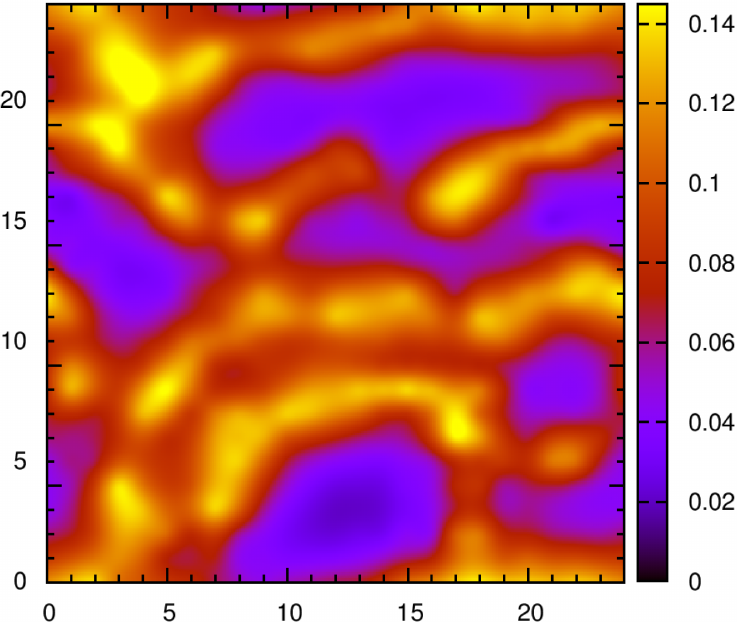}
    \put(-3,59){{\bf a}}
 \put(16,65){\large Order parameter}
  \put(22,71){\Large $\boldsymbol{x=1/10}$}
 \end{overpic}
\hspace{.5cm}
\begin{overpic}[scale=1,unit=1mm]{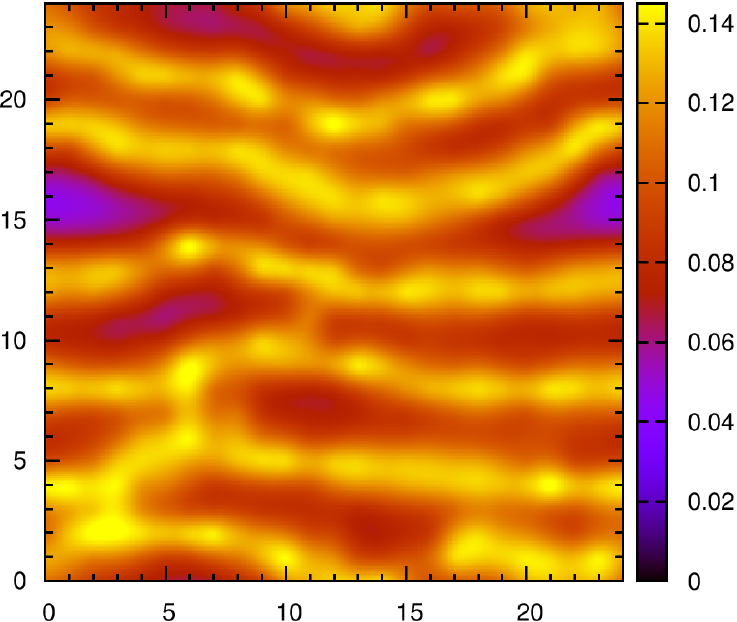}
    \put(-3,59){{\bf d}}
 \put(16,65){\large Order parameter}
  \put(22,71){\Large $\boldsymbol{x=1/8}$}
 \end{overpic}
\vspace{6mm}\\
 \begin{overpic}[scale=1,unit=1mm]{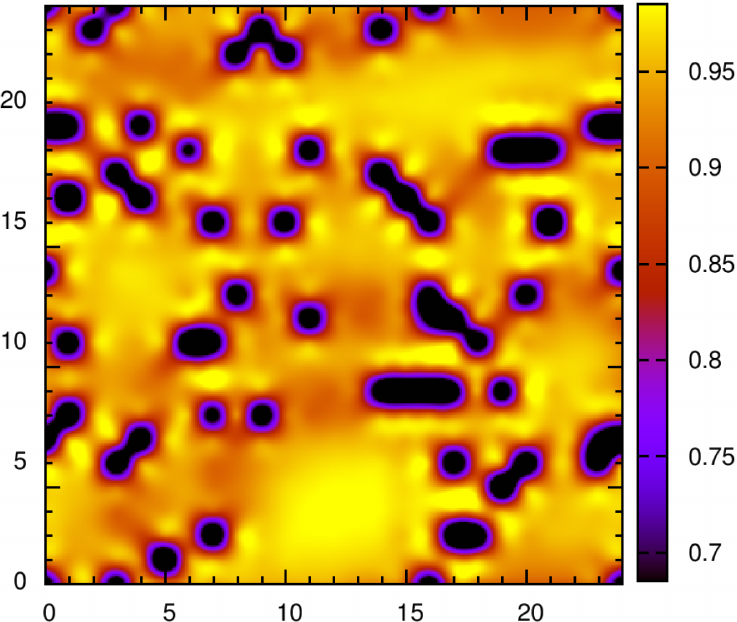}
    \put(-3,59){{\bf b}}
 \put(15,65){\large Electron density}
 \end{overpic}
\hspace{.5cm}
 \begin{overpic}[scale=1,unit=1mm]{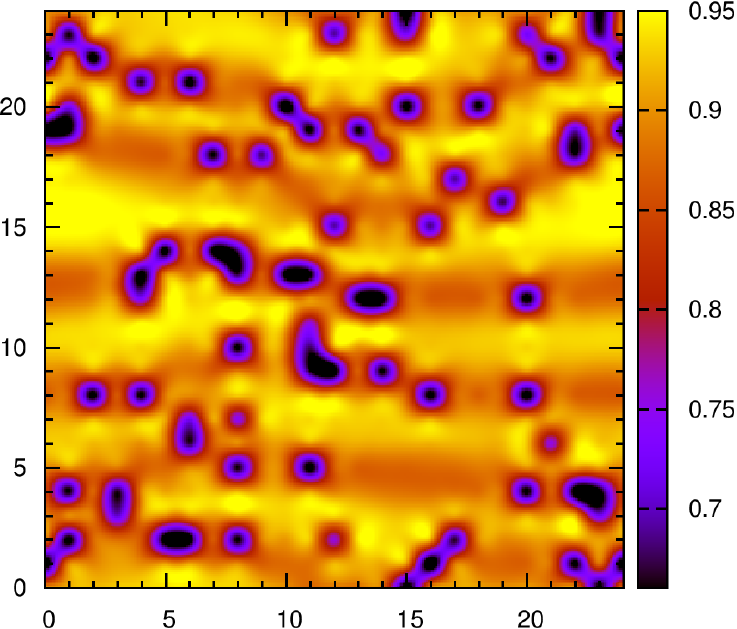}
    \put(-3,59){{\bf e}}
 \put(16,65){\large Electron density}
 \end{overpic}
\vspace{6mm}\\
 \begin{overpic}[scale=1,unit=1mm]{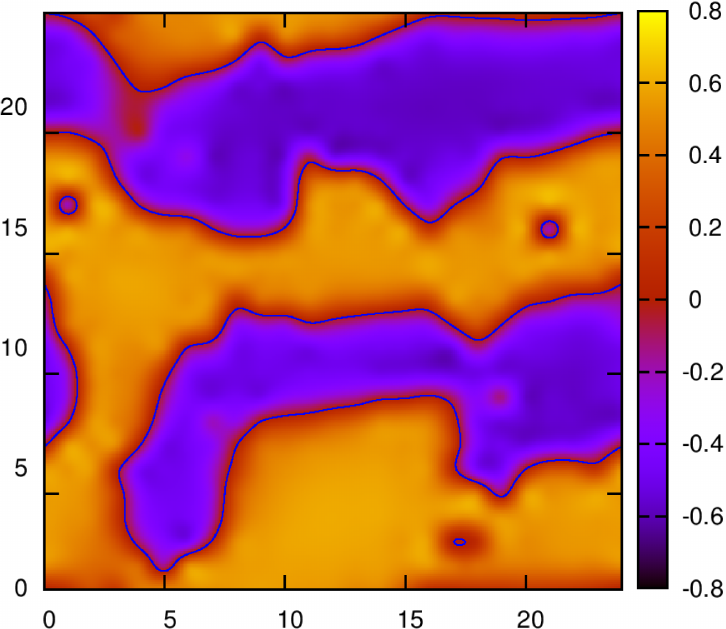}
    \put(-3,59){{\bf c}}
 \put(8,65){\large Staggered magnetization}
 \end{overpic}
\hspace{.5cm}
 \begin{overpic}[scale=1,unit=1mm]{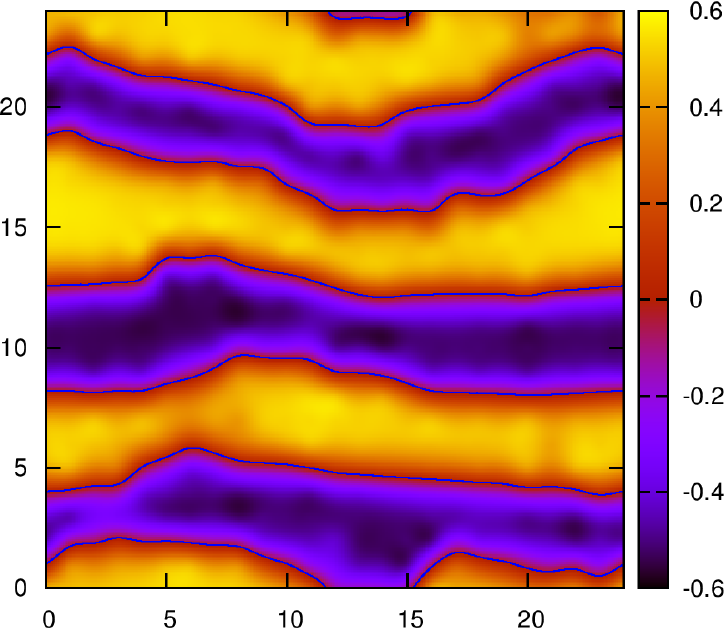}
    \put(-3,59){{\bf f}}
 \put(8,65){\large Staggered magnetization}
 \end{overpic}
\caption{\setlength{\baselineskip}{5mm} {\bf \boldmath Emergence and stabilization of stripes 
in real-space quantities of $d$-wave superconductors by a shift of hole doping from 
1/10 to 1/8 ($n_{\rm imp} = x, U = 3.3\,t$).} 
For this set of parameters density modulations emerge close to $x=1/10$ doping 
($V^{\rm imp} = 1.3t$) in the presence of strong dopant disorder and become unidirectional 
by a change of hole doping towards 1/8 ($V^{\rm imp} = 0.9t$).
(a), (d) SC order parameter $\Delta_i^d$, 
(b), (e) electron density, (c), (f) staggered magnetization 
(blue lines mark the zero-crossing) for $T=0.025\,t, t'=-0.4\,t, V = 1.6\,t$. } 
\label{fig:doping_dep}
\end{figure}

\begin{figure}[p!]
\begin{overpic}[scale=1,unit=1mm]{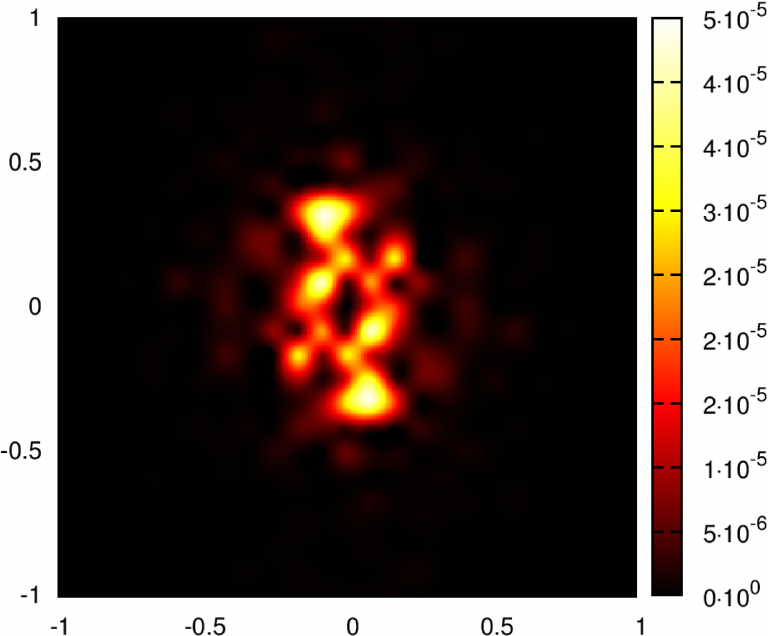}
    \put(-3,59){{\bf a}}
 \put(16,65){\large Order parameter}
  \put(22,71){\Large $\boldsymbol{x=1/10}$}
 \end{overpic}
\hspace{.5cm}
\begin{overpic}[scale=1,unit=1mm]{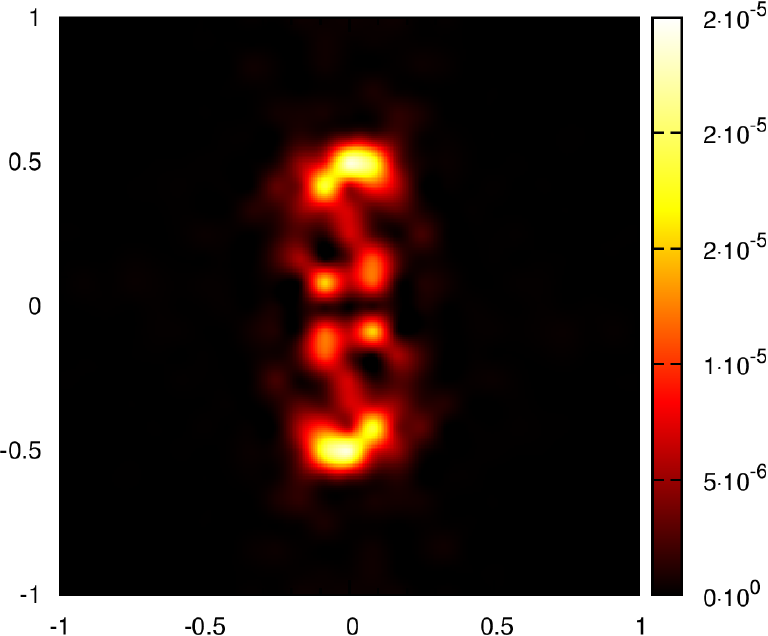}
    \put(-3,59){{\bf d}}
 \put(16,65){\large Order parameter}
  \put(22,71){\Large $\boldsymbol{x=1/8}$}
 \end{overpic}
\vspace{6mm}\\
 \begin{overpic}[scale=1,unit=1mm]{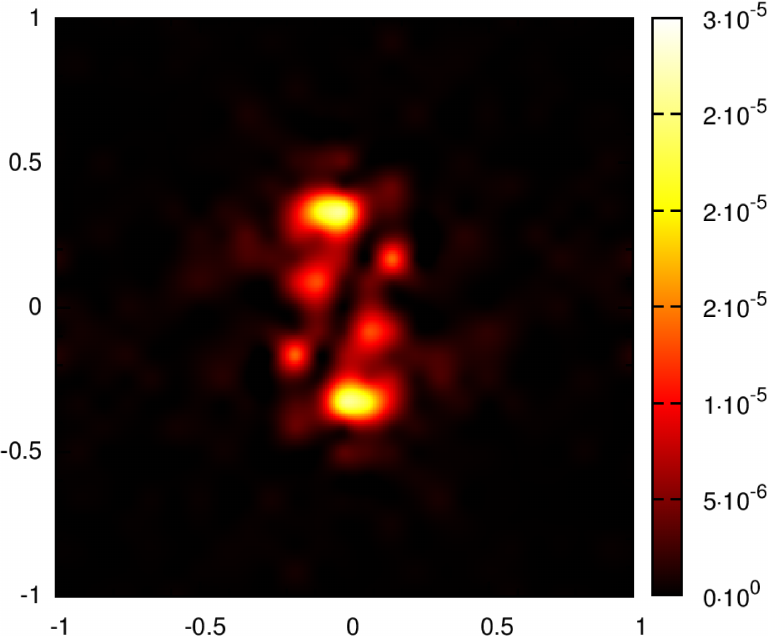}
    \put(-3,59){{\bf b}}
 \put(15,65){\large Electron density}
 \end{overpic}
\hspace{.5cm}
 \begin{overpic}[scale=1,unit=1mm]{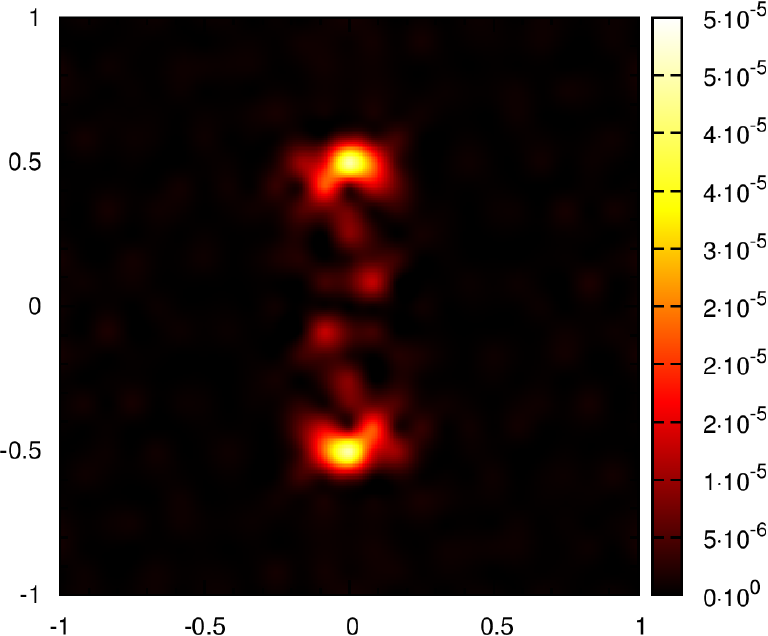}
    \put(-3,59){{\bf e}}
 \put(16,65){\large Electron density}
 \end{overpic}
\vspace{6mm}\\
 \begin{overpic}[scale=1,unit=1mm]{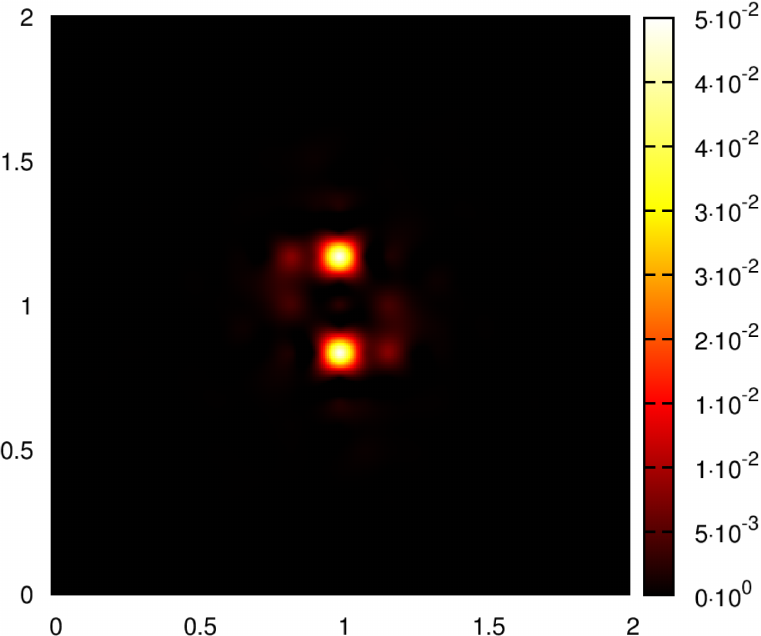}
    \put(-3,59){{\bf c}}
 \put(8,65){\large Staggered magnetization}
 \end{overpic}
\hspace{.5cm}
 \begin{overpic}[scale=1,unit=1mm]{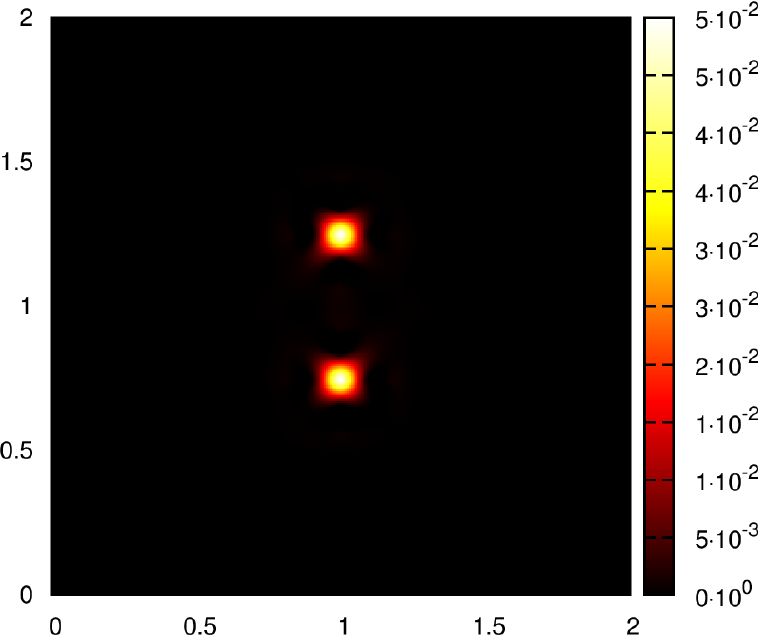}
    \put(-3,59){{\bf f}}
 \put(8,65){\large Staggered magnetization}
 \end{overpic}
\caption{\setlength{\baselineskip}{5mm} {\bf \boldmath Fourier transformed 
SC order parameter (a), (c), electron density (b), (e), and 
magnetic structure factor (c), (f) for 1/10 (a)--(c) and 
1/8 (d)--(f) hole doping.} 
The dopants' impurity potential is $V_{\rm imp} = 0.9\,t$ at $x=1/8$ and 
$V_{\rm imp} = 1.3\,t$ at $x=1/10$. Note that the impurity sites are not considered 
in the FT electron density (b), (e) as described in 
Sec.~\ref{sec:mom_space_quant}. 
Parameters were fixed to $n_{\rm dop} = x, U = 3.3\,t,  T=0.025\,t, t'=-0.4\,t, V = 1.6\,t$. 
Horizontal and vertical axes are given in units of $\pi/a$.} 
\label{fig:doping_dep_struc}
\end{figure}

For the modeling of cuprates, such as  BSCCO or those based on the parent 
compound La$_2$CuO$_4$,  
their dopant disorder has to be taken into account. 
In the following we consider explicitly 
intrinsic disorder 
and compare our results with 
experimental data in zero and finite magnetic field.  
For modeling dopant disorder we set 
the impurity concentration equal to the hole doping 
$ n^{\rm imp} = x$. 
Moreover, we assume that the dopants' 
impurity potential decreases with increasing hole doping due to enhanced screening 
\cite{andersen:2007}. 
In Fig.~\ref{fig:doping_dep} and Fig.~\ref{fig:doping_dep_struc} 
we show exemplary results for two specific impurity configurations 
which exhibit the characteristic behavior of the disordered stripe phase. We have 
verified that other impurity configurations yield density waves with the same wavelength 
as discussed here. 
For a Coloumb repulsion $U = 3.3\,t$ we find 
quasi-unidirectional stripes at $x=1/8$, slightly deformed by the presence of the non-magnetic 
impurities 
(see Fig.~\ref{fig:doping_dep}~(d)-(f)). 
Hole-rich paths (see Fig.~\ref{fig:doping_dep}~(e)) coincide with anti-phase domain walls, seen in 
the staggered magnetization profile in Fig.~\ref{fig:doping_dep}~(f), and with stripes of strong 
superconductivity as observed in the SC order parameter  Fig.~\ref{fig:doping_dep}~(d). 
Although the suppression of the electron density on the impurity sites dominates the 
charge pattern as seen in  Fig.~\ref{fig:doping_dep}~(e), closer inspection reveals the existence of 
hole-rich channels (colored orange in Fig.~\ref{fig:doping_dep}~(e)), 
where $\langle n_i \rangle$ is reduced compared to the average electron density. 
Additionally, the lines of zero staggered magnetization (blue lines in Fig.~\ref{fig:doping_dep}~(f)), 
which coincide with the maxima of the SC order parameter, also serve as a guide 
to the eye. The $d$-wave order parameter (Fig.~\ref{fig:doping_dep}~(d)) varies in the strong-disorder limit 
with much larger amplitudes as compared to the previous cases, but it never changes 
sign. 

From the analysis of different disorder configurations we conclude that the hole-rich paths emerge only close to $x=1/8$ hole doping and 
sharpen upon 
approaching $x=1/8$. 
Simultaneously the density 
modulations as a whole become quasi unidirectional. 
We find that, by shifting the hole doping 
towards 1/8, the system is driven into the stripe state. 
Moreover, we observe that the wavelength of the density modulation 
decreases
by enhancing the hole doping from $x=1/10$ to $1/8$ (cf 
Fig.~\ref{fig:doping_dep_struc}). 
This characteristic, which is also observed in 
neutron scattering experiments \cite{yamada:1998,fujita:2002}, 
is not found in the impurity-free systems. 
Experimentally, SDWs with wavelengths
$\lambda_{\rm SDW}(x = 1/10) = 10a$ and 
$\lambda_{\rm SDW}(x = 1/8) = 8a$  
were inferred from the incommensurabilities. 
While this model calculation predicts exactly the same wavelength 
for the SDW at $x=1/8$, the wavelength 
at $x = 1/10$ is slightly larger, that is 
$\lambda_{\rm SDW} = 12a$, than in the experiment. 
Note that a decrease in hole density either 
reduces the amplitudes of the density waves or enhances the wavelength. 
The latter is valid in this model calculation.
Strongly disordered systems support 
the former, where the random impurity sites facilitate the setup 
of additional hole channels.  
Disorder allows the 
stripe pattern to adjust more easily to the average hole density, whereas in the clean 
case, the stripe pattern is tied to the underlying lattice.

All real-space quantities 
verify horizontally (or vertically) oriented stripes at $x=1/8$. 
The periodicity of these patterns is extracted from the Fourier transformed (FT) 
quantities as shown in Fig.~\ref{fig:doping_dep_struc}. 
Intriguingly, the FT data in the strong disorder regime for $x=1/8$ (Fig.~\ref{fig:doping_dep_struc}~(d)-(f)) 
peak in general at the same 
wavevectors as for the perfectly unidirectional stripes discussed for clean systems. 
The magnetic structure factor (Fig.~\ref{fig:doping_dep_struc}~(f)) 
shows two dominating peaks at the  
incommensurate wavevectors ${\bf q}_m = \frac{2\pi}{a} (1/2, 1/2\pm\epsilon)$ with incommensurability $\epsilon = 1/8$. 
The FT electron density (Fig.~\ref{fig:doping_dep_struc}~(e))
and the FT SC order parameter (Fig.~\ref{fig:doping_dep_struc}~(d))
exhibit similar patterns with 
two broad maxima around wavevectors $\q_{c/p} \simeq 2\pi/a\,(0, \pm \delta)$ with $\delta = 1/4$. 
The main difference to the clean stripe solutions is that the peaks in 
Fig.~\ref{fig:doping_dep_struc}~(d), (e) are broadened, and 
a low-intensity substructure is observable. Obviously, the slight deviation from the 
characteristics of perfect unidirectional stripes is caused by the finite impurity 
concentration.
We infer that 
coexisting mdSC, CDW, and SDW with wavelength $\lambda_{\rm mdSC} = \lambda_{\rm CDW} = 4a$ 
and $\lambda_{\rm SDW} = 8a$ persist at $x=1/8$ even 
in the regime of strong disorder. 
The wavelengths of the density modulations change with doping 
as is recognized by comparison of 1/10 and 1/8 hole doping. 
In Fig.~\ref{fig:doping_dep_struc}~(a)--(c) the FT SC order parameter, the FT electron density, 
and the magnetic 
structure factor are displayed for $x=1/10$. The maxima deviate clearly from those at $x=1/8$. 
At 
$x=1/10$ the density modulations oscillate with wavelengths 
$\lambda_{\rm SDW} \simeq 12a$ and $\lambda_{\rm mdSC} \simeq 6a = \lambda_{\rm CDW}$. 
In contrast to {\it weakly} disordered systems the incommensurabilities $\epsilon$ and $\delta$ 
in the strong disorder limit are sensitive to doping and increase with growing hole concentration. 

\begin{figure}[t!]
\hspace{.1cm}
\begin{overpic}[scale=1,unit=1mm]{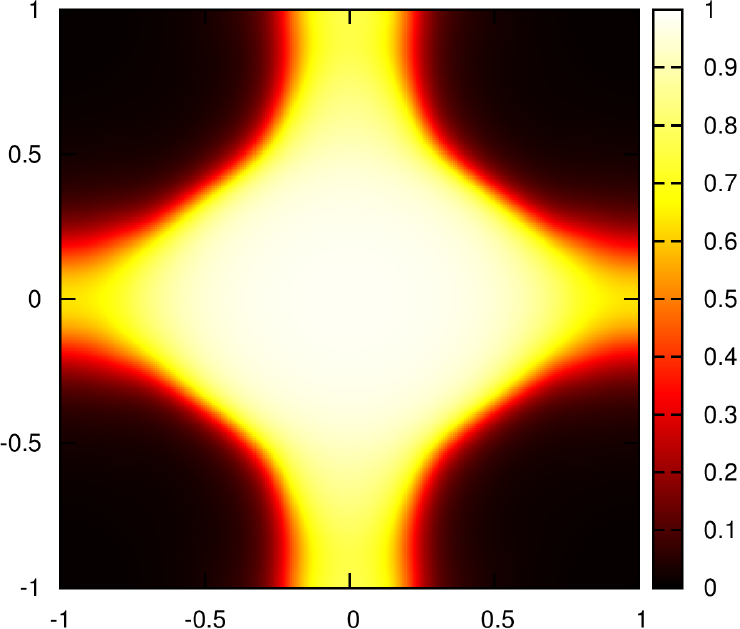}
    \put(-3,59){{\bf a}}
 \end{overpic}
\hspace{.6cm}
\begin{overpic}[scale=1,unit=1mm]{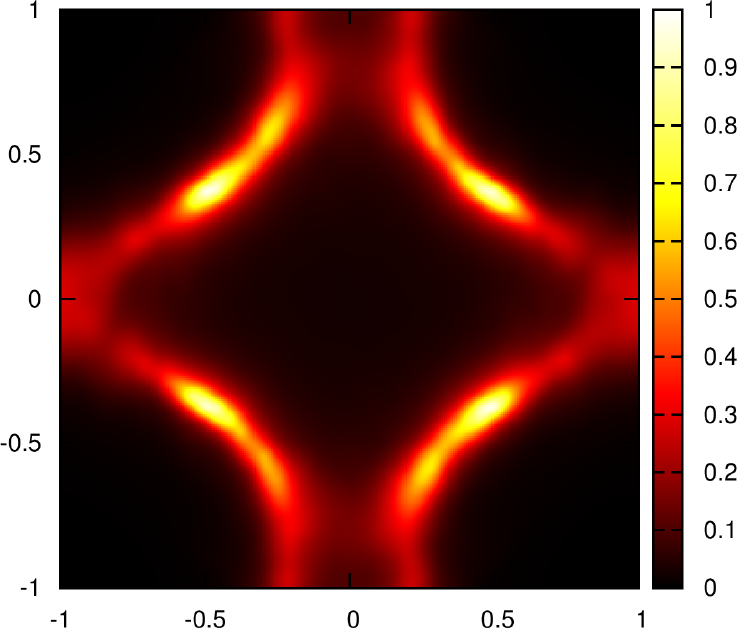}
    \put(-3,59){{\bf b}}
 \end{overpic}
\vspace{1em}
\hspace{.1cm}
 \begin{overpic}[scale=1,unit=1mm]{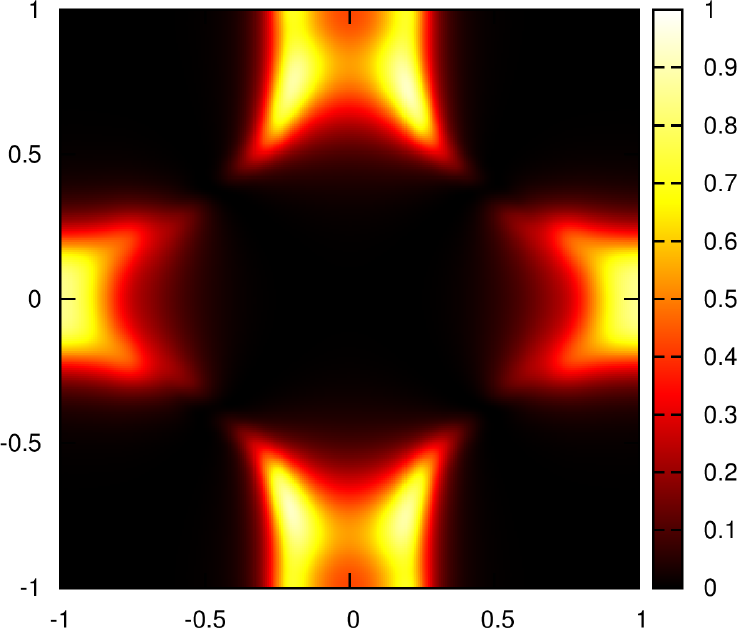}
    \put(-3,59){{\bf c}}
 \end{overpic}
\hspace{.6cm}
 \begin{overpic}[scale=1,unit=1mm]{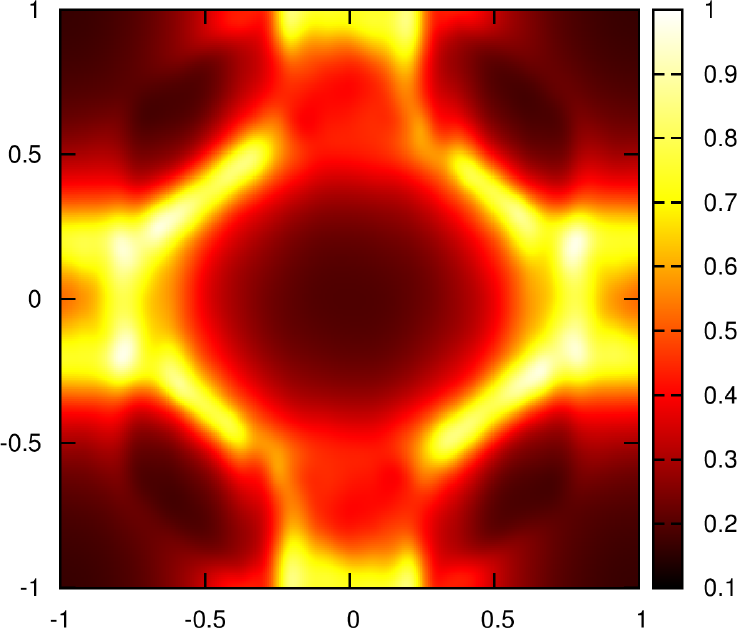}
    \put(-3,59){{\bf d}}
 \end{overpic}
\caption{\setlength{\baselineskip}{5mm} {\bf \boldmath Momentum-space quantities in the 
strong disorder limit.} 
a) Momentum distribution $n(\k)$, 
(b) spectral density $A(\k)$, 
c) pair density $P(\k)$, and (d) spin density $\rho_s(\k)$ for
 $T=0.025\,t, t'=-0.4\,t, V = 1.6\,t, U = 3.3\,t, x=1/8 = n_{\rm dop}, V^{\rm dop} = 0.9\,t$. 
Horizontal and vertical axes are given in units of $\pi/a$.} 
\label{fig:disorder_qspace}
\end{figure}

The momentum space quantities 
$n(\k)$, $A(\k)$, $P(\k)$, and $\rho_s(\k)$ (see Fig.~\ref{fig:disorder_qspace}) 
in the strong disorder limit
resemble the results obtained for 
the impurity-free stripe solutions (see Fig.~\ref{fig:impfree_qspace}).  
Disorder slightly enhances
the anisotropy, best visible in $\rho_s(\k)$. 
Importantly, in the disordered system continuous Fermi arcs appear 
in the near nodal direction 
(see Fig.~\ref{fig:disorder_qspace} (b)). 
This agrees with ARPES measurements in LBCO \cite{valla:2006}, 
which has a strong dopant disorder. Due to the impurity potentials, through which 
the hole channels run, the CDW is stronger here as compared to the previously 
discussed cases, which results in the redistribution of spectral weight to the nodal 
directions \cite{jiang:2009}. 

Altogether, we find that a SDW, a CDW, and a mdSC coexist at $x=1/8$ hole doping 
in strongly 
disordered systems such as dopant disordered LSCO. The density-modulations are pinned 
and stabilized, just as for a single impurity or a vortex, yet slightly deformed 
by the strong disorder. 
The doping dependence of the incommensurabilities $\epsilon \propto x$ and 
$\delta \propto x$ is similar to those observed in experiments \cite{yamada:1998,fujita:2002}.

\subsubsection{Density of states}

\begin{figure}[t!]
\vskip2mm
\centering
\vskip2mm
\hspace{-.6cm}
\begin{overpic}[width=9cm]{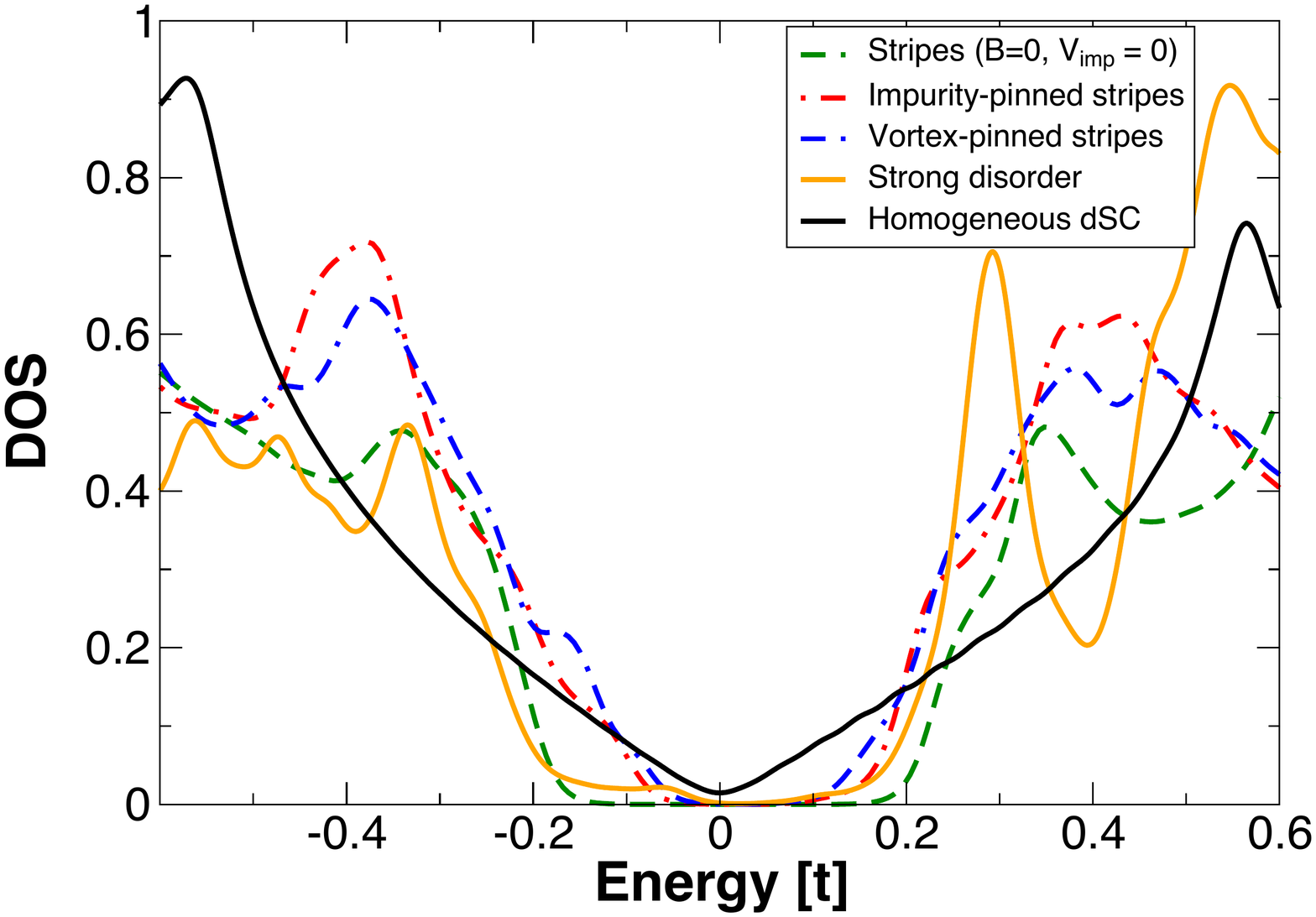}
\end{overpic}
\caption{\setlength{\baselineskip}{5mm}{\bf DOS in $d$-wave superconductors in the 
presence of 
stripes.}  Results for the data sets used above for impurity-free, 
impurity-pinned, vortex-pinned, and disorder-stabilized stripe solutions for $x=1/8$. 
The black curve corresponds to the DOS 
of a homogeneous $d$-wave superconductor.
}\label{fig:dos_stripes}
\end{figure}

Characteristically all the identified stripe solutions exhibit 
a full gap in the density of states (DOS) (see Fig.~\ref{fig:dos_stripes}). 
For comparison also the 
typical v-shaped gap of a homogeneous $d$-wave 
superconductor is shown in Fig.~\ref{fig:dos_stripes}. 
In the presence of stripes 
a full gap opens around the Fermi energy. 
The full gap is a generic property of stripes in $d$-wave superconductors 
which persists also in the strong disorder limit. 
The full gap originates from 
a finite extended $s$-wave contribution in the striped $d$-wave superconductor 
and the SDW gap present in the strongly magnetized stripes. 
In a homogeneous $d$-wave superconductor 
the bond order parameter $\Delta_{ij}$ changes sign between 
horizontal and vertical bonds which reflects the $d$-wave symmetry 
of the energy gap $\Delta_{\k}$. 
Thus the summation of $\Delta_{ij}$ over the vertical {\it and} horizontal bonds 
connecting to a given lattice site gives zero in a homogeneous $d$SC. 
Due to the strong anisotropy of the stripe solutions, however, 
this summation yields a finite value here, which characterizes extended $s$-wave 
superconductors. 
In a $d$SC coexisting with a SDW the gap is generically not centered around the Fermi 
energy, which is the reason why the full gap is asymmetric in all stripe solutions shown 
in Fig.~\ref{fig:dos_stripes}. In addition, particle-hole symmetry is already broken by 
the constant random impurity potentials with $V_{\rm imp} > 0$. 
Experimentally, particle-hole anisotropy was observed recently by 
angle-resolved photoemission spectroscopy (ARPES) in the pseudogap 
phase of Bi-2201 \cite{hashimoto:2010} 
which is attributed to competing orders  
and contrasted to homogeneous superconductivity. 
The appearance of a full gap was also found by Loder {\it et al.} 
\cite{loder:2010}.  
Obviously, the gap is largest in the unperturbed stripe solution ($B = 0, V_{\rm imp} = 0$). 
This is due to the fact that with the impurity- and field-induced bound states, 
spectral weight is shifted into the energy gap.

\subsection{Impurities in the $\bm V$-Model}\label{sec:V}

\begin{figure}[t!]
\begin{overpic}[width=0.485\columnwidth]{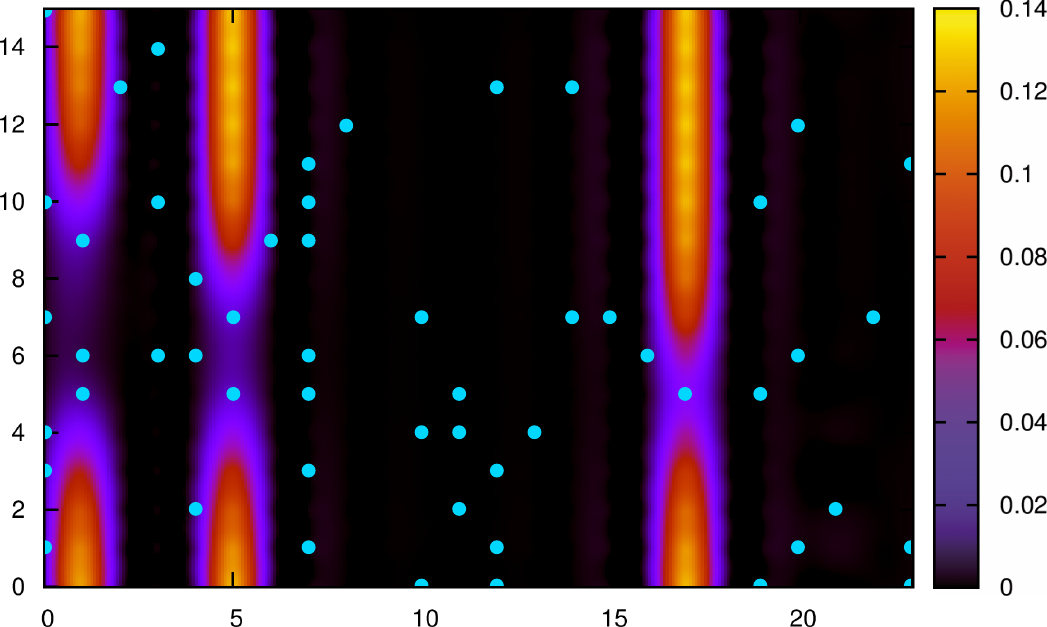}
\put(-9,126){{\bf a}}
\end{overpic}
\hspace{.3cm}
\begin{overpic}[width=0.485\columnwidth]{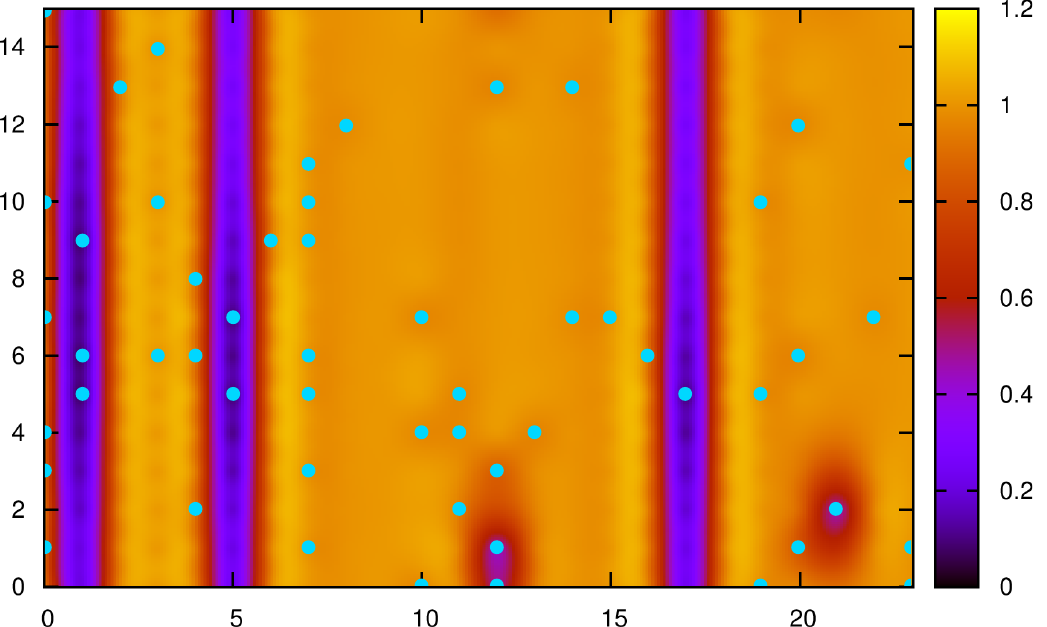}
\put(-9,126){{\bf b}}
\end{overpic}
\vskip6mm
\begin{overpic}[width=0.485\columnwidth]{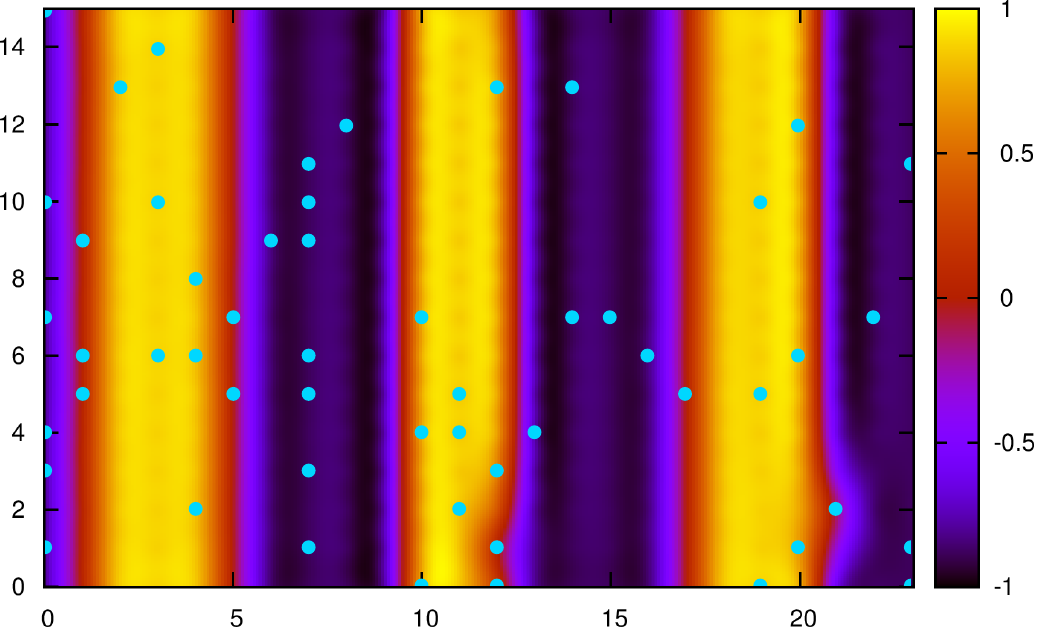}
\put(-9,126){{\bf c}}
\end{overpic}
\hspace{.3cm}
\begin{overpic}[width=0.485\columnwidth]{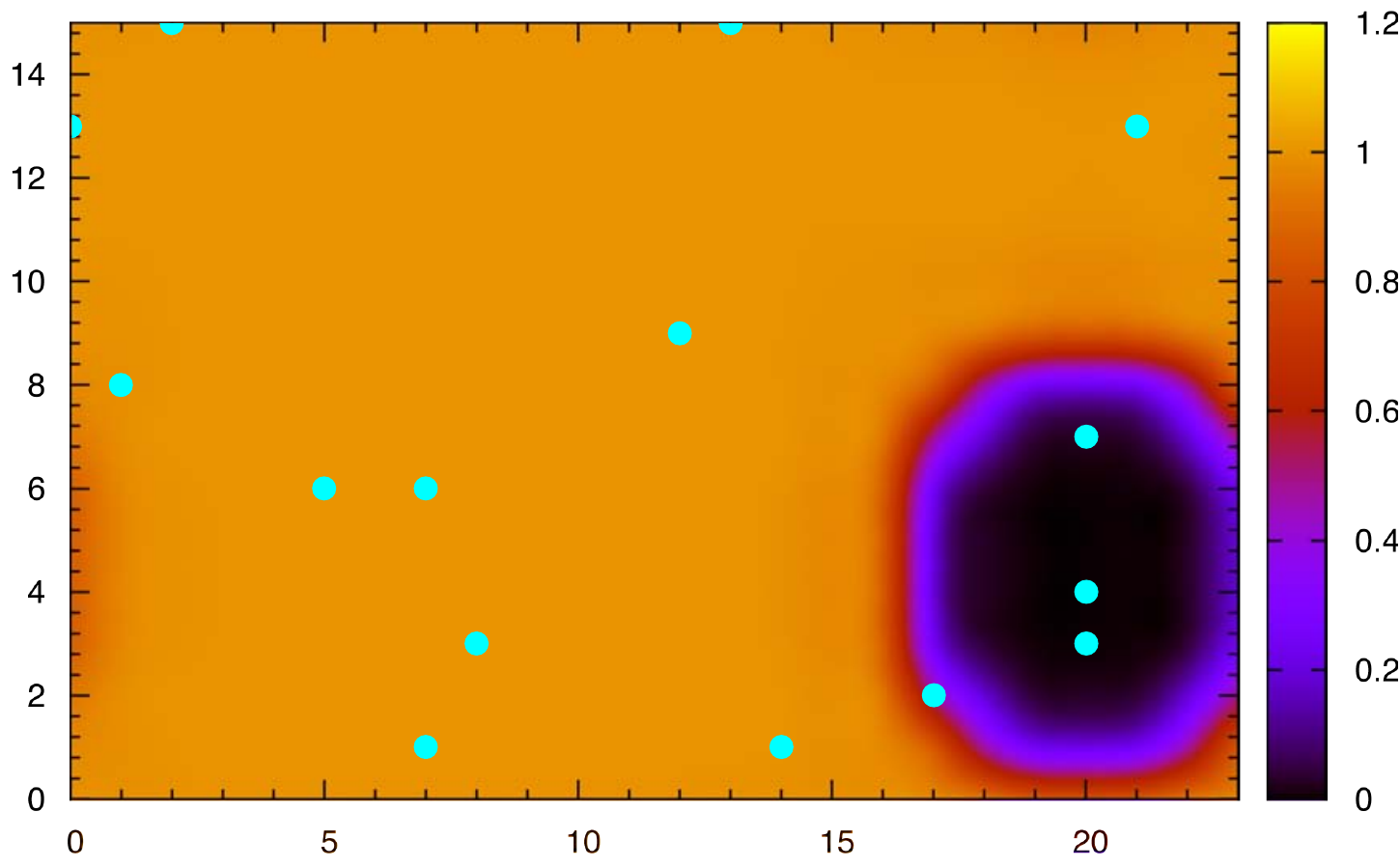}
\put(-9,126){{\bf d}}
\end{overpic}
\caption{\setlength{\baselineskip}{5mm} {\bf \boldmath Stripe patterns in the disordered 
$\bm V$-model.} 
(a)--(c) Weak impurity potentials $V^{\rm imp} = 0.9t$ ($x = 1/8 = n_{\rm dop}$) 
and (d) strong impurity potentials $V^{\rm imp} = 2.0t$ ($n_{\rm dop} = 4\%$). 
(a) $d$-wave order parameter $\Delta^d_i$. (b) Electron density. (c) Staggered magnetization. 
(d) Electron density exhibiting global phase seperation for strong impurity potentials. 
The turquoise dots indicate the positions of the impurities.
Parameters were fixed to $T=0.01\,t, t'=-0.4\,t, V = 1\,t$.} 
\label{fig:V}
\end{figure}

The $V$-model presented by the Hamiltonian Eq.~\eqref{eq:v} 
was introduced in Ref.~\cite{loder:2010} as an alternative description 
for the coexistence 
of superconductivity with magnetic order. 
In the homogeneous case, the groundstate solution of the $V$-model for $x = 1/8$ is 
a mdSC state similar to Fig.~\ref{fig:impfree}(a), (c). 
The main difference is that the AF stripes are nearly maximally magnetized, 
i.e. the SDW oscillates with an amplitude $\approx \mu_\text{B}$, while 
$\Delta^d_i$ goes to zero at the center 
of the AF stripes without changing sign. 
Thus magnetic order is dominating 
in the $V$-model, 
whereas superconductivity is reduced to straight, 
quasi one-dimensional lines on the domain walls between the AF stripes. 
This mdSC solution is degenerate with a ``pure'' PDW, which features a periodic sign 
change in $\Delta^d_i$. 
Both, the ``pure'' PDW and the mdSC have 
identical pair densities $P^2(\k)$ (c.f.~Ref.~\cite{loder:2010}). 

In the strong disorder limit the $U$- and the $V$-model show a severely different behavior. 
The solutions of the $V$-model can be divided into two regimes. If the scattering 
strength $V^{\rm imp}$ of the impurities is weak (i.e. $V^{\rm imp}\lesssim t$), the AF stripes remain 
straight, as shown in Fig.~\ref{fig:V}. 
The magnetic energy gain is dominant and prevents the superconducting stripes 
to wind around the impurities, which would allow to gain further condensation energy. 
Moreover, impurities in the straight one-dimensional SC channels strongly suppress 
superconductivity. This effect culminates in the disappearance of entire 
SC stripes, if the impurity density in their near vicinity becomes too large (Fig.~\ref{fig:V}~(a)).  
One of the neighboring AF domains (see Fig.~\ref{fig:V}~(c)) spreads over 
such a metallic line expelling the holes, which collect in the 
remaining superconducting stripes (Fig.~\ref{fig:V}~(b)). 
In this situation the AF domain-wall boundaries are shifted locally, according to the 
impurities' positions, 
but the AF spin stripes remain largely straight (Fig.~\ref{fig:V}~(c)).
The antiferromagnetic stripes are almost half filled 
and all sites within the AF stripes are nearly fully polarized, 
in contrast to the small amplitude modulations seen in 
Fig.~\ref{fig:doping_dep} for the $U$-model.  
Thus the superconducting stripes in the disordered 
$V$-model have a
filling below 1/4 because the holes collect 
mainly in the remaining SC stripes. Similar to the SDW, also the CDW in the $V$-model modulates 
with much larger amplitudes as compared to the $U$-model.

If the impurity potentials $V^{\rm imp}$ become larger than $t$, we expect that 
the impurity-pinning of stripes 
is energetically favorable as long as a mdSC state prevails, similar to the $U$-model 
(c.f. Fig.~\ref{fig:doping_dep}(a)--(c)). 
However, we observe an overall disappearance of superconductivity 
accompanied by a 
global phase separation. 
All holes accumulate where the impurity 
concentration is largest and antiferromagnetism weakest, while the surrounding 
region
turns into a half-filled AF insulator (Fig.~\ref{fig:V}~(d)). 
The same occurs if weak impurities are numerous enough to destroy 
sufficiently many metallic 
(superconducting) lines so that the remaining ones 
cannot accommodate 
all the holes in the 
system. This behavior is characteristic for the strong-coupling limit described by the 
$V$-model: if the effectively one-dimensional striped system is forced to 
become two-dimensional because of disorder, the system divides into half-filled 
and empty regions featuring global phase separation.

\subsection{Discussion of the experimental observations in different cuprates}\label{sec:dis}

We found that the $U$-model allows for two-dimensional superconductivity in coexistence with a stripe order that adapts flexibly to disorder, while the solutions of the $V$-model remain quasi one-dimensional. Materials in which three-dimensional 
superconductivity coexists with stripe order, such as 
LBCO~\cite{huecker:2011} 
or the rare-earth doped cuprate La$_{2-x-y}$Nd$_{y}$Sr$_{x}$CuO$_4$ with $x\lesssim0.12$ 
and $y=0.2$ \cite{buechner:1994,tranquada:1995}, are certainly 
not as one-dimensional as the 
solution of the $V$-model. 
On the other hand the $V$-model reproduces the quasi 
one-dimensional characteristics of Nd-doped LSCO around $y=0.4$, 
quite well \cite{loder:2010}, where the material becomes more anisotropic with increasing doping. 
As argued in Ref.~\cite{loder:2011}, the models described by 
${\cal H}_U$ and ${\cal H}_V$ are the weak and strong coupling limits of a $t$--$J$ like 
model, and the cuprates are most likely found somewhere in the intermediate 
coupling regime.

The static, unidirectional CDWs and SDWs observed in LBCO and LNSCO \cite{fujita:2004} are 
often ascribed to the structural phase transition towards 
the LTT phase, where the LTT specific buckling pattern of the CuO$_6$ 
octahedra induces an $x$-$y$ anisotropy \cite{kampf:2001}. 
Our results imply that the strong 
dopant disorder, which is present in these substances, further stabilizes the stripe state. 
The results within the $U$-model in the strong disorder limit reproduce typical properties 
of these substances well, such as the Fermi arc reconstruction in LBCO \cite{valla:2006} or 
LSCO \cite{razzoli:2010} 
and the doping dependence of the wavelength of the SDW as inferred from neutron scattering 
experiments in LSCO \cite{yamada:1998,fujita:2002}. 
 
Though spin stripes are observed in LSCO at $x = 1/8$, 
an ordering of the charges has so far {\it not} been detected by neutron scattering experiments 
\cite{suzuki_lsco:1998, kimura:1999}. 
As LSCO does not exhibit an LTT phase 
we suggest that the experimentally observed spin stripes 
\cite{suzuki_lsco:1998} are induced by the dopant disorder. The expected concomitant 
charge stripes are not found by neutron scattering experiments possibly because the charge 
modulations are weak and may therefore be concealed 
by the reduction of the electron density below the dopant sites. 

YBCO, which has neither an effective dopant disorder nor a LTT phase, exhibits 
dynamic instead of static stripes 
\cite{stock:2004, vojta:2010}. 
This agrees with the results of the $U$-model  
for an impurity-free system as discussed. 
In the presence of a 
magnetic field charge stripes become static, while spin stripes are absent 
\cite{wu_ybco:2011}. In this case the magnetic field induced charge stripes 
erase superconductivity. 
We find a similar behavior in the $U$-model when we turn 
off superconductivity deliberately. The magnetization 
is greatly suppressed, while a dominant charge order survives.

\section{Summary}\label{sec:sum}

The models we investigated in this article allow to explain a wide range of features of striped cuprate 
superconductors, which were found experimentally.  
Notably, the material specific characteristics of stripe phases in various 
cuprates can be modelled within the $U$-model. 
On the other hand, we found that 
AF correlations in the $V$-model tend to be stronger and the superconducting stripes 
are one-dimensional which reproduces some characteristics of Nd-doped LSCO 
quite well.

Specifically, we found  
the coexistence of PDW's, SDW's, and CDW's with wavelengths 
$\lambda_{\rm CDW} = \lambda_{\rm PDW} = 4a = 1/2\,\lambda_{\rm SDW}$ in a $d$-wave 
superconductor for sufficiently large $U$ in the $U$-model. 
Impurity-free, (single) impurity-pinned, and vortex-pinned {\it straight} stripe 
solutions exist over a broad doping range in the $U$-model. Disorder dissolves the strict 
one-dimensional stripe order into meandering stripe-like patterns. 
Only close to 
$x=1/8$ hole doping, which is ideal for stripe formation, 
a unidirectional stripe order reemerges in strongly disordered systems. 

In the $V$-model stripes 
are rigid and remain virtually 
undistorted in the presence of weak disorder. 
Superconducting stripes react very sensitively to disorder 
which may even remove some of the superconducting channels. 
The remaining superconducting stripes exhibit strictly 
one-dimensional superconductivity 
appearing between straight AF stripes. Strong disorder, however, leads to global phase separation 
into dominantly antiferromagnetic regions with non-magnetic puddles absorbing the holes. 

The overall identical characteristics of the impurity-, vortex-pinned, and 
(strong) disorder-pinned stripes indicate
that the stripes are not generated by impurities or fields. 
Instead, stripe states, which are energetically close to 
homogeneous solutions in unperturbed systems, are pinned and stabilized 
by inhomogeneities which are connected to charge redistributions. 

A generic feature of striped $d$-wave superconductors in both the $U$- 
and the $V$-model is the opening of a full gap, 
induced by an extended $s$-wave contribution. 
Without local probes for  
the superconducting order parameter it is not 
clear if pair-density modulations accompany the charge- and spin stripe order. .

\vspace{5mm}
\noindent
{\bf Acknowledgements:}
\noindent We thank Siegfried Graser for helpful discussions. 
This work was supported by the Deutsche Forschungsgemeinschaft through TRR 80.

\end{document}